There's something I would like to understand.
And I don't think anyone can explain. …
There's your life. You begin it, feeling that it's
something so precious and rare, so beautiful
that's like a sacred treasure. Now it's over,
and it doesn't make any difference to anyone,
and it isn't that they are indifferent, it's just
they don't know, they don't know what it means.

Ayn Rand
We the living

# On the "barcode" functionality of the DNA,

## or

# The phenomenon of Life in the physical Universe

Simon Berkovich
Department of Computer Science
The George Washington University
Washington, DC 20052, USA
Telephone:  (202) 994-8248
Fax:         (202) 994-4875
E-mail:       berkov@seas.gwu.edu

# ABSTRACT

The information contained in the genome is insufficient for the control of organism development. Thus, the whereabouts of actual operational directives and workings of the genome remain obscure. In this work, it is suggested that the genome information plays a role of a "barcode". The DNA structure presents a pseudo-random number (PRN) with classification tags, so organisms are characterized by DNA as library books are characterized by catalogue numbers. Elaboration of the "barcode" interpretation of DNA implicates the infrastructure of the physical Universe as a seat of biological information processing. Thanks to the PRNs provided by DNA, biological objects can share these facilities in the Code Division Multiple Access (CDMA) mode, similarly to cellular phone communications. Figuratively speaking, populations of biological objects in the physical Universe can be seen as a community of users on the Internet with a wireless CDMA access. The phenomenon of Life as a collective information processing activity has little to do with physics and is to be treated with the methodology of engineering design. The concept of the "barcode" functionality of DNA confronts the descriptive scientific doctrines with a unique operational scheme of biological information control. Recognition of this concept would require sacrificing the worldview of contemporary cosmology.

# Contents

**Preamble**

A limited practical success can be reached with a mistaken scientific theory. Thus, the construction and exploitation of steam engines was perfectly compatible with an innocuous idea that heat is a liquid. And this example is not a rare exception. However, undermining rational thought with erroneous believes eventually lead to a dead-end. Particularly misleading is a belief in causal connections in lack of sufficient information. Suppose that someone has been presented a still image and then has been shown moving pictures claiming that these moving pictures originate from this still image. For a reasonable person such a connection can be only nominal, like a connection between an advertisement on a box and a full-fledged movie inside this box.

Radical scientific advancements need new methodology. Thus, the development of calculus was initiated by Newton's mechanics. But, "the great era of mathematical physics is now over" (Berlinsky, 2001). Understanding of complex behavior must rely on algorithmic approach. The progress of biology needs a new methodology – the methodology of engineering design of information processing systems. The engineering design methodology has broader facilities than regular application of computer modeling. This methodology addresses fine construction points that in computer programming can be omitted.

The breakthrough comes with the development of a cellular automaton model, which portrays the physical world as a gigantic information processing machine. The phenomenon of Life is a network activity where biological objects are seen as a community of users on the "Internet" of the physical Universe. In our time, when everything goes digital with a distasteful trivialization and surprising effectiveness a noted remark by Edward Teller should be paraphrased: "Technology of to-day is science of to-morrow".

The essence of the suggested organization of biological information processing can be illuminated by considering the complexity of the "black box" problem in airplanes for possible recovery of crash related information. A simpler and more effective solution presents a design where the required information is collected outside of the flying airplane through communications.

The suggested concept is very general and could be opposed from all directions. The best available science interprets the genome as a set of entities "responsible for" specific traits (gene for intelligence, gene for the shape of the face, gene for a particular disease etc). This interpretation is "partly correct" and can be compared to a description of a movie story by characters shown in a still picture. Centuries ago, Molière had ridiculed such an approach with his notorious "vis dormitiva" (opium is effective because it has "soporific power"). Common descriptions of biological processes can not withstand a routine engineering analysis. Functioning of living systems has little to do with physics and chemistry. It is a problem of information control.

The "barcode" interpretation of the DNA furnishes a natural explanation to two surprising facts: paradox N - how an organism can be built from an information deficient genome and paradox C - why more complex organisms have less complex genomes. The DNA molecules get control signals through communication, so shorter structures acquire an operational edge. In simple words, DNA is a label name and to a certain extent a shorter name is an advantage.

## 1. Introductory remarks

"One of the profoundest enigmas of nature is the contrast of dead and living matter" (Weyl, 1949). Envisioning living organisms as machines governed by laws of physics and chemistry raises the sacramental question of whether the "living matter" posses some properties which are not inherent to the "dead matter". Anyway, why and how does the change in the behavior of dead and living matter occur so abruptly?

The core of the enigma of living matter lies in the origin of information control. The problem is how biological objects acquire guidance through the life cycle as they emerge from and degrade into non-existence. In this paper, the organization of biological information processing is approached purely in terms of "engineering design". The everlasting debates on delicate points of the origin and true meaning of Life present a separate issue.

The contemporary science firmly rests upon the conviction that Life is a mere by-product on top of the material processes. The life cycle of a biological organism is considered as a sequence of transitions from one molecular configuration to another. This outlook is problematic in many respects. Particularly confusing is the fact that the structural complexity of the genome is insufficient for organism development.

The deficit of the genome information is supposed to be compensated via "interaction with the environment". In our suggestion, the information in the genome does not play its traditional role of data or instructions. Instead, the information contained in the DNA macromolecules presents a pseudo-random number (PRN), like, for example, a barcode. The structure of the DNA is not a primitive carrier of fixed information resources but an identification key that ensures a unique specification of an organism within a broad taxonomy. Using the DNA "barcode" as a key for wireless communications by means of the Code Division Multiple Access (CDMA) technique biological organisms acquire access to incomparably richer information processing facilities.

How can resemblance of a child to the father be conveyed through the information deficient genome? The "barcode" interpretation of DNA simply indicates that functioning of biological objects cannot be based on material configurations. To understand the phenomenon of Life it is necessary to consider informational infrastructure underlying the material world. The idea that Life is associated with immaterial information processes in the Universe had been around, in one form or another, through the whole history of human civilization. In the suggested construction this idea is linked to a cellular automaton model of the physical world. The basic pillars of this model - hardware componentry and software architecture - are outlined in Appendices A and B.

The Appendices A and B consider general problems: how the "barcode" functionality affects the view on the physical Universe and what changes does it incur in the computational scheme of biological information processing. The main body of the paper concentrates on issues of the "barcode" functionality of the DNA that are of immediate bio-medical concern.

## 2. The groundwork theses

1. The "barcode" functionality of the DNA gives the genome an operational meaning.

2. The amount of information involved in biological information processing is enormous.

Complications in fundamental biology simply reflect the fact that information associated with Life and Mind overwhelms the diversification of the material world.

3. Biology must comply with "the basic law of requisite variety" which says that achieving of appropriate selection "is absolutely dependent on the processing of at least that quantity of information. Future work must respect this law, or be marked as futile even before it has been started"(Ashby, 1962).

4. Investigation of biological information processing must rely on the methodology of engineering design. There is a limit of complexity that anything can sustain - the design must be done with "economy and elegance".

5. Pure technicalities in the implementation of information processes in the physical Universe should not be mixed up with the philosophical and metaphysical questions of higher meaning.

6. Problems such as how memories are stored in the brain are not likely to be affected by the discovery of the final theory in physics (Weinberg, 1992). Understanding of biological information processing will come with the re-vitalization of the concept of ether (Wilczek, 1999; Davis, 2001). Foundations of physics have to be revised in conjunction with the involvement of information:

   1). The material structure of the DNA molecules does not contain enough variety to serve as a repository of control directives for living organisms (Claverie, 2001).

   2). The existing picture of information pathways in the physical Universe is incomplete. Information impact of quantum entanglement behind material processes spreads at least $10^7$ faster than light (Seife, 2000).

   3). Modern cosmology does not care about information structures in the Universe. Instead, the bulk of the Universe (95%) is supposed to be filled with an unstructured stuff of "dark matter" and "dark energy" (Cowen, 2001).

7. A. Einstein stated: "Someday we'll understand the whole thing as one single marvelous vision that will seem so overwhelmingly simple and beautiful that we will all say to each other -- Oh, how could we have been so stupid so long? How could it have been otherwise?" Ironically, Einstein's concept of general relativity is the primary barrier on the way of the information dominant Universe.

8. The major instrument in the advancement of knowledge is *Experimentum Crucis* - a crucial experiment that demonstrates a clean fact negating a contender theory. Confirming evidences do not assure logical correctness of a scientific theory - the moment of truth comes through a negation. The decisiveness of an *Experimentum Crucis* in natural science can be compared to that of a *counter-example* in mathematics and an *alibi* in jurisprudence.

9. Implications of the "barcode" functionality of the DNA hit the basis of science and thus may lead to a cultural shock. A comprehensive model of the Universe necessarily involves an excess of miscellaneous minute particulars that are difficult and tedious to scrutinize.

10. The convincing power of the suggested idea is going to be increased by virtue of some compelling circumstances:

   1). Intensive investigations of the pseudo-random composition of the genome would reveal very limited knowledge about the essence of Life, if at all.

   2). The suggested concept predicts startling destructive effects of cross-talks and interference in clones. With upcoming mass production of clones such vital outcomes could not pass unnoticed.

   3). A number of organism's disorders, like "mad cow disease", may originate from malfunctioning of the underlying information processing mechanism below the level of conventional physiology.

## 3. The invisible biological computing

The statement that biological objects must be backed up with intensive computing is a truism, but informational pathways in the physical world remains inscrutable. In a wide range of speculations, the organization of biological information processing is somehow related to a mystifying holistic process. Thus, in (Talbot, 1991) biological processes are associated with a holographic mechanism of the Universe. The suggested scheme of biological information processing gets hardware support from the holographic mechanism of the cellular automaton model of the physical world (Appendix A).

Workings of a holographic mechanism need a source of coherent reference, recording, and reconstructing waves. The important feature of the suggested cellular automaton model is a continual generation of synchronizations and desynchronizations that spread through the whole Universe at the frequency of ~$10^{11}$ Hz. These wave-like activities provide reference waves for recordings and reconstructions. The internal rhythm at the frequency ~$10^{11}$ Hz permeating the whole Universe serves as the clock pulse generator for all biological systems. The presence of a clock generator at ~$10^{11}$ Hz is associated with a class of biological effects that develop without apparent reasons under subtle influences of microwaves at this frequency. "From the experiments on millimeter waves… it can be concluded with reasonable confidence that these waves cause effects that can be understood neither in terms of heating nor through direct action of the electric fields of the waves. It follows that the electromagnetic wave acts as a trigger to events for which the biological system is already prepared" (Fröhlich, 1980).

There is a great confusion in comparison of the brain and the computer. The number of switching events per second apparently shows at least a 10,000 times advantage of the computer over the brain (see, e.g., Hillis, 1985): "Thus the sheer computational power of the computer should be much greater than that of the human. Yet we know the reality to be just the reverse. Where did the calculation go wrong?"

The slow neural circuitry cannot supply a substantial computational power. The extremely high information processing capabilities of the brain must be related to a different hardware mechanism. Resolution of the mystery of the organization of the brain should count on extracorporeal placement of human memory (Berkovich, 1993). Information processing in the brain is based on the holographic memory of the Universe operating at the characteristic frequency of ~$10^{11}$ Hz. This organization employs an unconventional computational model using fast memory and slow processing elements. The *Experimentum Crucis* for the suggested concept develops the paradoxical observations of Gregory, 1959 (Appendix B).

The underlying informational infrastructure of the physical Universe - the place of biological information - is a cellular automaton network of processing nodes. Material world does not correspond to the "hardware" of the network, as a superficial consideration would rush to suggest. Instead, elementary constituents of the material world are patterns of synchronization activities. These synchronization patterns, having the corresponding properties of elementary particles, can be combined together according the conventional laws of physics and chemistry to present atoms, molecules, and the bulk matter (Appendix A).

The population of biological objects using external information processing resources of the physical Universe can be seen as a community of users on the Internet. In this analogy, biological objects can be considered as "migrating software agents". The genome provides an "access card" to network resources. Human brain is a "terminal" attached to the network, not a standalone computer, in famous poetical words: "No man is an island". There is a resemblance in workings of human minds and searching engines.

Spreading the precious contents of human memory over the whole network rather than storing it in one vulnerable location is an obvious design advantage. The major concern about the ability of living systems to withstand the trend of destruction imposed by the principle of the entropy increase (Schrödinger, 1992) goes away as the stabilization of large molecular complexes is maintained by exterior control signals.

## 4. The DNA as the key to informational resources of the physical Universe

The central problem for the suggested organization of biological information processing is how to apportion the information processing facilities of the physical Universe among the zillions of living systems. The resolution of this problem is associated with the "barcode" workings of the DNA (Berkovich, 1999a and 1999b).

A particular instance of the genome creates a specific pattern of conformational oscillations in the chromosomes. This pattern modulates the holographic waves in the informational infrastructure of the physical world. At the same time, chromosomes are susceptible to these patterns of oscillations, so DNA can extract them from the correspondingly modulated incoming waves. The set of modulation patterns of the DNA in different chromosomes represents a pseudo-random number (PRN) characterizing a specific genome composition. This PRN serves as a "barcode" distinguishing a given organism in the whole the system of other living beings.

Biological objects having different PRNs modulate their information transmissions in different ways. The communication bandwidth of the physical Universe can be shared using the Code Division Multiple Access (CDMA) technique. The same principle implements shared access to the content-addressable holographic memory. The CDMA approach to bandwidth allocation problem is used in the cellular phone technology. Figuratively speaking, one can say that the DNA is not a "blueprint" for organism construction, but rather a "cellular phone" through which the directives for this construction are obtained.

The answer to the sacramental question what causes the drastic differences between the "living matter" and "dead matter" lies in the size of the molecules involved. A short PRN associated with small material formations can pick up only noisy background transpiring as long-range quantum correlations. A lengthy PRN can sustain a robust information exchange, so macromolecules of the DNA serve as CDMA transmitters and receivers. Also, the macromolecules of DNA act as microtransducers materializing control signals into purposeful biological events. When a modulation pattern of feeble incoming signals matches the pattern of the conformational oscillations along the large DNA structure the impact of these signals can be amplified and turned into a tangible mechanical action.

Individual biological objects can share the information processing resources of the physical Universe – the communication bandwidth and the content-addressable holographic storage – because the PRN of their DNA structure present what is called spreading sequence (see e.g. Stallings, 2001). Effectually, a long spreading sequence encodes the transmission of one bit of information from a corresponding source. When the signal is received, the transmitted bit of information is extracted by using the same spreading sequence to remove the encoding. The transmission rate of spreading sequence is substantially higher than the transmission rate of actual data.

Several "engineering" advantages are gained from this kind of organization of biological information processing:

- the transmitted information gets protection from various kinds of distortion;
- only the owner of a unique spreading sequence is in command of its transmissions;
- many biological objects can independently use the available bandwidth and storage with very little interference.

To constitute a workable system the spreading sequences have to be orthogonal in the sense that actions of different sequences should result in mutual cancellation. Two general categories of spreading sequences have been used: pseudo-random sequences and orthogonal codes. As an example of the later, consider a set of four vectors presented by +1 and −1: (+1,+1,−1,−1), (+1,−1,+1,−1), (+1,−1, −1,+1), and (−1,−1,−1,−1). Any pair of these vectors is orthogonal in the sense that their dot product is 0. Now, consider the case of pseudo-random sequences of a very large size N. Thanks to the Law of Large Numbers different long pseudo-random sequences will be approximately orthogonal with the accuracy of $1/\sqrt{N}$. For different organisms to be informationally separated it is vital that their DNA structures would be substantially dissimilar. An occasional closeness of DNA structures of different organisms can produce some cross-correlation resulting in inconclusive manifestations of extra-sensory perception.

The CDMA communication systems can support users at different rates of data transmission. This can be done by using spreading codes of different lengths while still maintaining their orthogonality: shorter spreading sequences have to be orthogonal to arbitrary segments of equal length within larger spreading sequences. Fortunately, this requirement is fulfilled automatically for the set of spreading sequences in the form of large pseudo-random numbers. The time of transmission of a data bit encoded by a spreading sequence is proportional to the length of this sequence. It is very important to note that users with shorter spreading sequences operate at higher transmission rates.

## 5. Paradoxes of the genome composition

Examination of genome characteristics reveals strange facts contradicting to ordinary expectations. All these considerations favor the idea that the genome information is rather an identification label than a container of directives.

1. From the point of view of information theory the genome is a text composed of the four DNA nucleotides letters: A, T, C, G. To decipher a hidden meaning of a text it is useful to start with estimate of its statistical entropy. Entropy analysis of the DNA structure does not reveal a significant departure from randomness; this indicates a characteristic property of an identification label, not of a meaningful text.

2. Vast stretches of the DNA in a genome serve no purpose and thus called "junk DNA". The operative genes constitute only 3% of the genome. For a set of construction directives such a non-operative redundancy is wasteful. The barcode interpretation makes a great sense of this mysterious circumstance: the "junk" part of the DNA structure simply presents a unique identification number while the operative genes provide classification tags with individual and species specific information.

3. It is reasonable to expect that more complex biological objects would require a genome of a larger size. However, this is not the case. Thus, the amount of the DNA in some species of amoebae is about 30 times as large as that in humans. There is a perplexing circumstance with plants that have more DNA than some animals. So, it appears that the amount of information in a DNA structure has little to do with its operational capabilities.
Moreover, longer identification code means a lower transmission rate of control signals in cell communications. Thus, a smaller DNA is more beneficial for cell control than a larger DNA.

4. There are many examples of a disproportional strength of genes. Thus, there is a very small difference in genes between human and monkeys - less than 2%. Also, the genome of mice is quite similar to the genome of humans. So, how can it happen that some genes become so powerful in effect? In the barcode interpretation of the DNA such a question would not appear because in a label all letters are "equal". The difference in the "strength" of identification tags is the difference between the "strength" of the objects, which they describe. Analogously, in a toy store a single difference in a barcode can distinguish between a "fly" and an "elephant".

5. Individual variations in the genome appear in a systematic way as minute differences between genes - single-nucleotide polymorphism (SNP). In the presented concept, the small distinctions between the genomes of different organisms is not an incidental circumstance but a critical factor required for the biological individuality of organisms.

6. From the conventional standpoint it is unclear why every cell of a developing multicellular organism has to carry the whole set of chromosomes. But the fact that every cell carries a complete set of genetic information is essential for organism control. With the barcode functionality of the DNA the significance of this fact is clear: carrying a complete set of DNA molecules, every cell gets access to the same communication facilities. As a result a multicellular organism becomes a coherent system whose elements can work under a centralized control.

7. If not for the "barcode" interpretation of the DNA it would be unclear how the same gene can play substantially different roles in different genome contexts.

## 6. The vital role of biological individuality

Cells with the same DNA have a common key to information processing resources of the physical Universe. It is a distinctive mark of the biological individuality of an organism, which brings its cells under a unified control. The unique DNA structure secures for all the cells of an organism the same slice of the information processing resources of the physical Universe. Thus, the biological individuality of organisms is a necessary factor of their existence.

Most explicitly the biological individuality of organisms manifests itself in workings of the immune system. The distinctive characterization of an organism provides protection from invading pathogenic viruses and bacteria. This system is responsible for recognition of the pathogens and producing an army of specific antibodies. The immune system performs a number of complex tasks: (1) regulating immune responses to discriminate between "self" and "not-self" antigens, (2) encoding the whole repertoire of about 1 million possible antibodies, (3) arranging immunological memory that can store a particular history of organism's reactions over many decades. With the suggested organization of biological information processing the intricate properties of the immune system are clearly elucidated.

## 7. Randomization of DNA structures

The validity of the presented concept hinges upon the possibility of newly created organisms to acquire random information. In multi-cellular organisms this acquisition occurs at the stage of meiosis through the process of recombination. In a chaotic exchange of parts of paternal chromosomes, chances that some of the PRNs in different organisms would be close are very low. So, normally organisms are protected from extraneous access. A rare possibility of close PRNs shows up as a slight sporadic malfunction of the system. On the other hand, in single cell organisms, like bacteria, DNA generated by mere replications are identical. Therefore, populations of bacteria can create interrelated communicating networks (see, e.g. Sonea, 1988).

Randomization of DNA structures is a remarkable feature of biological progressions: "pieces of DNA are continually being separated and brought together in new combinations"; it appears that "the major consequence of sex is to make genetic recombination possible" (Smith,1986).

Mere replications of DNA in microorganisms are not capable to provide a great diversity of PRNs. Also, organisms created by parthenogenesis have limited possibilities to obtain differences in PRNs, so they depend on occasional sexual reproduction. Importantly, "species which wholly abandon sex are short-lived on an evolutionary time-scale" (Smith,1986). The reason for this is determined by limited information processing capabilities of organism's account associated with a given PRN. Combination of sexual reproduction and replication in certain insects, like bees and ants, may sustain the biological individuality only for the colony but not for its members.

The important issue of biological individuality of monozygotic twins has not been overlooked in our analysis. It has been indicated in (Berkovich, 1999b) that monozygotic twins can almost certainly enjoy biological individuality since they have low chances (1 in 500,000) to get completely identical PRNs (assuming that the formation of monozygotic twins is determined by a portion of the chromosome set as suggested in Berkovich and Bloom, 1984).

## 8. The life cycle

The development of a new organism starts with opening "an account" on the "Internet of the Physical Universe" using the DNA in a zygote as an identification key. With this key each organism gets an access to a "slice" of the holographic storage of the physical Universe. Any operation within an organism produces a Read-Write transaction at this account. The content-addressable access prevents information in a holographic storage from being erased selectively.

The arrow of biological time is irreversible because information produced and recorded in the course of organism development cannot be deleted. With the accumulation of information at organism's account the control signals associated with this account progressively lose the precision in the retrieved information. There is a limit on the amount of information that can be effectively utilized for control purposes. This factor establishes an upper bound on the lifespan of all biological organisms.

The operational capacities of a biological memory system are restricted by the diversity of content-addressable encoding. Roughly speaking, with an access register of W bits it is not possible to write down in a content-addressable storage more than $2^W$ different words. The chromosome structure of every living being exactly presets the range of codes to exercise their life cycle.

Reducing the information influx to organism's account can increase longevity. The increase of longevity could occur irrespective of the means by which this reduction is achieved. It has been noticed that alike organisms of thinner and smaller constitution live longer. The expansion of livespan due to the caloric restriction sustains the same explanation as it spares organism's account by diminishing the intensity of biochemical transactions. The longevity might be also increased due to deprivation of information from sensory inputs (Apfeld and Kenyon, 1999).

Accumulation of non-erasable information utilized in life cycles of old organisms affect the development of next generations of organisms. The phenomenon of Life is a developing collective process with an intrinsic property of inheritance of somatic changes. Thus, the necessity to change is an indispensable part of the existence of biological systems. The suggested mechanism results in a specific Lamarckian type of evolution. This mechanism explains a profound philosophical controversy of how random perturbations, which are a major destructive factor for physical systems, appear as the sole constructive determinant for biological systems (Monod, 1972).

The cardinal points of the organism life cycle are recapitulated in Fig. 1

## 9. Two scenarios of catastrophic interruption of control flows

Many malfunctionings of a human organism considered as "molecular" diseases are in fact "information" diseases. An information irregularity could occur as a result of disruption in the supply of control signals to the cells of an organism. This disruption may cause catastrophic effects as it touches the basic control signals going through the DNA. The flow of control signals can be impaired in two different ways: when the DNA key is fractured or when a flawless key is restrained by an extrinsic impact. These possibilities bring together under a unified operational scheme two scenarios of cell destruction in different pathological cases: genome dysfunctions, like AIDS, and neurological disorders, like prions diseases.

Both scenarios result in the same outcome - disruption of cell control signals. Lack of control signals at the element level implies a slow but tenacious degradation of cells. Cell control signals acquired by means of the DNA are fundamental for the organization of Life, so their disruption is almost impossible to withstand. The situation with cancerogenesis presents a different way of organism destruction. The case of cancer may be simpler because it deals with a loss of control at the level of cell populations rather than with a damage of control inputs at the level of individual elements.

### 9.1. Impact of gene substitution

It is taken for granted that an ailing organism can be cured if a correct gene substitutes a defective one in cell chromosomes. An organism with a "good" gene instead of a "bad" gene is supposed to become healthier like a car after one of its damaged parts is replaced. Such a "repair" philosophy, however, may be inadequate when applied to substitution of components in information processing systems.

The notion of "good" or "bad" genes is instrumental at the stage of the very beginning of the organism development when its "information account" is not yet fully formed. For a mature organism any "new" gene, in general, may be always bad. A new gene changes the "key" to the account and thus deprives the cells of established operative control signals for the evolved organism. Deficit of these signals may result in simultaneous failure of the cells over the whole organism. This scheme of massive collapse of cells corresponds to an unfortunate picture in some attempts to use gene substitution as a therapeutic procedure.

Contrariwise, the gene substitution technique applied to cancerous cells might be more fulfilling. According to the presented concept, a population of cancerous cells may subdue from getting any kind of "new" genes.

Retroviruses HIV transcribing their RNA information in the form of DNA in the T-cells of the immune system produce a gene substitution effect, which invalidates the "barcode" key of these cells. Thus, cells of the immune system begin to degrade. This concept suggests different models for the development of AIDS in infants and adults. Untreated infants born with HIV may accommodate extraneous information in their DNA "barcode" and to a certain extent develop like "regular" organisms, possibly with some genetic anomalies. In adults, HIV disrupts the cell control information for an already established account. So, for infants with an innate HIV the pathological picture of AIDS should unfold differently in a less devastating way than for adults.

## 9.2. Impact of non-genetic materials

The mechanism how non-genetic materials, like proteins – prions, can destroy cell constructions is not clear (Prusiner,1995). Basically, two circumstances are surprising: how agents that do not replicate can affect growing cell populations and why a severe damage comes with particular conformational variations. An explanation can be given using the suggested barcode interpretation of the DNA.

Proteins harbored inside a cell are not involved in the transmission of the cell control information, but they can influence this process indirectly by modulating conformational oscillations of the DNA. In the case of gene substitution, the disruption in cell communications happens because cells use a wrong "barcode" key, in the case of an extraneous intervention, cell communications are performed with a right "barcode" key but are restrained by conformational influences of contaminating proteins.

The disruption of cell communications by external agents depends on the kinetics of the interplay between the processes of growth and contamination. Slowly dividing neural cells are more sensitive to the intervention of non-genetic materials as shown by the following simplified model. Suppose that cell contamination is going at a constant speed and takes time – $T_P$ to reach the critical level; suppose that cell divisions occur at a steady rate with time interval between cell divisions being $T_D$. Let $p_n$ present relative levels of cell contamination after $n^{th}$ division ( $n = 0, 1, 2…$). Since non-genetic materials do not replicate, as a result of cell division the level of contamination halves. This can be described by a recurrence equation with an initial value $p_0 = 0$:

$$p_n = p_{n-1} / 2 + T_D / T_P \qquad (1)$$

The solution of this equation is given by a geometric series which for sufficiently large n approaches 2 ($T_D / T_P$). The contamination process can reach the critical level, $p_n$ approaching 1, only if:

$$T_D > T_P / 2 \qquad (2)$$

This means that the destruction of cells under steady contamination by non-genetic material can occur when the time between cell divisions, $T_D$ , is sufficiently large. So, this destruction can happen for slow-dividing cells, like neural cells, but not for fast-dividing cells. This conclusion supports the idea suggested in (Prusiner,1995) that a similar prion type contamination may be also responsible for some neurodegenerative diseases where pathogenic agents are difficult to identify.

## 10. Experimentation with the genome cross-talks and interference

The major trait of the "barcode" functionality of DNA – shared access to information resources of the physical Universe - leads to a possibility of communication cross-talks among various biological objects having identical DNA structures. These cross-talks occur by communication through common memory, so, they are not fixed in time and can extend over indefinitely long periods. Higher organisms employ recombination of chromosomes to ensure randomization of their DNA keys. Still cross-talks of higher organisms cannot be excluded completely, but they are exceptional and their observations are inconclusive. Under these circumstances, the traditional scientific method aimed at establishing cause-effect relationships becomes ineffective. The interference of biological objects through common memory can spread over arbitrary distances in space and long delays in time, so finding a reasonable cause-effect relationship may be problematic.

Multi-cellular organisms with closest DNA structures are clones. In artificially produced clones the suspected cross-talks can be investigated in repetitive controlled conditions. A newly created clone does not open its own "account" at the "Internet" of the physical Universe but simply enters the already existing "account" of the donor. Thus, a clone inherits the age of the donor and, not surprisingly, undergoes premature aging and untimely death. Also, clones often fail to develop into normal organisms. But apparently cloning does not damage animals' genes since clones can produce healthy offsprings. This fact gets an excellent explanation in the suggested concept: the development of a clone as a new organism is hampered by the extraneous information already accumulated in the mature account of the donor, while the development of clones offsprings, which acquire a new randomized DNA key, starts afresh. Yet abnormalities in an individual clone may be furnished with some common explanations. The unexpected effects of cross-talks and interference can be revealed in observations of systems of clones.

The "account" corresponding to identical DNA keys of a clone and its donor will be under joint usage and thus will run out in a shorter period of time. This leads to an unusual hypothesis that cultivating a clone can shorten the lifespan of the clone donor. The effect may be increased in cultivating a multiplicity of clones from the same donor. No explanation of this anticipated result can be obtained from the standpoint of the conventional paradigm. The supposition that cultivating clones and donors in entire isolation can affect their lifespans emphatically highlights the essence of the suggested concept.

At the level of microorganisms with mere replication of DNA, the cross-talks of biological objects are more pronounced. To survive, the populations of microorganisms have to undergo continuous transformations to change DNA keys for switching from one used up "account" to another. As an example, one can turn attention to the emergent resistance of bacteria to antibiotics. Observations (Gilliver, 1999) show a buildup of resistance to antibiotics in the absence of a traceable exposure. It would be interesting to examine whether microorganisms with some acquired property being germinated in a carefully secluded location can indeed induce proliferation of microorganisms with a similar property on a global scale.

## 11. Conclusion

The idea that the DNA structure characterizes a biological object as a "barcode" makes perfect sense from the viewpoint systems design. The role of the DNA label in Nature is similar to that of a Social Security Number in the society – identification of individual objects. The unworkable "junk" DNA in the genome of more than 95% is useful for the purpose of identification.

The DNA structure does not contain the required operational information, so the directives for organism development have to come from outside. This points to the infrastructure of the material world as a source of control information for biological objects. Considering Life as a collective activity raises specific "engineering" problems of organization of a distributed multi-user environment. The key role in this organization is played by the DNA, whose pseudo-random structures provide orthogonal spreading sequences for sharing the bandwidth and storage of the Universe using the Code Division Multiple Access (CDMA) protocol. Shorter spreading sequences provide a higher transmission rate and thus are more efficient in elucidation of the surprising discovery that more complex organisms have simpler genomes.

The "barcode" interpretation of DNA signifies the triumph of the reductionism in the sense that functioning of biological objects can be ultimately reduced to elementary constructions of the material world as long as they are supported by the underlying infrastructure of the physical world. The suggested model for the phenomenon of Life in the physical Universe is conceptually very simple. It is instructive to quote Hawkins, 1988 that a complete fundamental theory "should in time be understandable in broad principle by everyone, not just a few scientists."

The suggested scheme of biological information processing relies on the design of the physical Universe as a gigantic cellular automaton network. Typically, a design of a system has to be supported by a huge documentation elaborating finest engineering details. The Appendices A and B can be treated as an outline of supportive "documentation" of "hardware" and "software" for materialization of living systems in the physical Universe. There is a number of fundamental questions about the properties of the physical world look very profound, mysterious, and impenetrable (see, e.g. Schewe and Stein, 2001). From the standpoint of the presented design of the Universe many of the features of the physical world appear obvious and lackluster. As J. M. Keynes said: "The difficulty lies, not in the new ideas, but in escaping from the old ones".

## 12. Acknowledgements

Scarce is the truth, but the supply was always in the excess of the demand. I am grateful to many of my devoted friends for their valuable insights and enthusiastic support of this long and thorny enterprise.

# LIFE  CYCLE

| | |
|---|---|
| ***MEIOSIS*** | RECOMBINATION |
| | CREATION OF PSEUDO-RANDOM  NUMBERS |
| | OF DNA SEQUENCES |
| | |
| ***FERTILIZATION*** | FORMATION OF A ZYGOTE |
| | ACQUIRING  BIOLOGICAL  INDIVIDUALITY |
| | GENOME  IS  A USER ID  FOR |
| | "OPENING  AN  ACCOUNT" |
| | |
| ***BIOLOGICAL  INDIDUALITY*** | THE NECESSITY OF THE UNIQUE IDENTIFICATION |
| | THE SPECIFICITY OF THE IMMUNE RESPONSE |
| | |
| ***ORGANISM  DEVELOPMENT*** | DIFFERENTIATION AND MORPHOGENESIS |
| | SENSING  GENERAL SPECIES-SPECIFIC CONTROL |
| | SIGNALS AND ACCUMULATING PRIVATE |
| | INFORMATION |
| | |
| ***GAINING  CONSCIOUSNESS*** | EXTRACTING  INNATE  TEMPLATES  OF THE MIND |
| | DEVELOPED PATTERNS APPEAR AS NEURON SPECIALIZATION |
| | |
| ***AGEING*** | ACCUMULATION OF  NON-ERASABLE INFORMATION |
| | IRREVERSIBILITY  OF  BIOLOGICAL  TIME |
| | INFORMATION POLLUTION IMPEDES CONTROL SIGNALS |
| | |
| ***DEATH*** | EXHAUSTING THE RANGE OF ENCODING FOR ACCESS KEY |
| | LIFE SPAN OF HUMANS  –  $3 * 10^{20}$  READ/WRITE OPERATIONS |
| | AT WORKING FREQUENCY  $10^{11}$ Hz |
| | |
| ***LIFE  AFTER  DEATH*** | THE  ACCOUNT  DETERMINED  BY THE GIVEN |
| | DNA STRUCTURE IS CLOSED |
| | INFORMATION ACCUMULATED BY A DEAD |
| | ORGANISM REMAINS IN THE  SYSTEM OF THE |
| | PHYSICAL UNIVERSE BUT CANNOT BE  ACCESSED |
| | WITHOUT A CORRESPONDING  PRN |
| | |
| ***THE CAUSE OF EVOLUTION*** | INFORMATION  ACCUMULATED  BY  AN  ORGANISM |
| | DURING  ITS  LIFE  CYCLE  AFFECTS  NEW  ORGANISMS |
| | THESE IMPACTS MAY BE REMOTE IN SPACE AND TIME |

Fig. 1

The mission of the genome

Dear reader or, or better still, dear lady reader, recall the brightfull, joyful eyes with which your child beams upon you when you bring him a new toy, and then let the physicist tell you that in reality nothing emerges from these eyes; in reality their only objectively detectable function is, continually to be hit by and to receive light quanta.  In reality! A strange reality! Something seems to be missing in it.

<div align="right">Erwin Schrödinger</div>

## Appendix A. Towards the concept of information dominant Universe

Fundamental physics, as understood nowadays, cannot continue to enjoy its eminence without discounting Life as a by-product on top of the material processes. The anticipated final theory in physics has nothing to contribute to the organization of biological systems. Trying to unify physics first and then seeking for an explanation of Life looks like an attempt to "cross a chasm in two small jumps."

Is the recourse to "dark energy" overpowering the Universe a scientific achievement or a misleading failure? The "barcode" interpretation of the DNA stretching outside for the source of control information is unyielding. Therefore, rejecting the idea of an infrastructure dominating the information processes in the physical Universe condemns the fundamental biological science to non-advancement.

The foundations of biology are at the mercy of intellectual processes beyond its reach. For the time being, the functionality of the genome is practiced in abstract way.  Without a clear operational scheme one just say that separate genes have been assigned to certain functions for which they somehow manage to get information from the "environment".

The current picture of the Universe prevents understanding of living systems on the basis of the "barcode" interpretation of the DNA. Anyway, modern cosmology is on the verge of restructuring because of its own internal inconsistencies. "While it may be the conventional wisdom that Einstein had finally got the workings of the universe right, the history of science strongly suggests that this is presumptuous and highly unlikely" (Petrosky, H., "To engineer is human. The role of failure in successful design", Vintage Books, New York, 1992).

Two things are necessary for a success of a new theory: a crucial experiment to reject the existing doctrine and a new model to collect the shattered facts.

## 1. *Experimentum crucis*  -  A Non-Dopplerian Ingredient in the CMB dipole

The whole Universe is pervaded with microwaves having a spectrum of a blackbody radiation at the temperature about 2.72 $^{\circ}$K. Our consideration of this process is radically different from the common view on its mechanism. According to the conventional doctrine the Cosmic Microwave Background (CMB) is a remnant of the radiation from matter left behind soon after the Big Bang. Thus, the CMB must have a perfect spherical symmetry except for an imprint of small fluctuations that had subsequently transpired in the clump structure of the Universe. Investigations of the CMB are in the mainstream of current astrophysical research.

Yet the observed distribution of CMB temperature shows a significant dipole deviation from uniformity - this distribution is an ellipsoid rather than a sphere. The immediate explanation is simple and natural: the dipole asymmetry in CMB is a result of the Doppler effect due to the global motion of the solar system.

Our consideration brings in an idea that the CMB distribution does not have the original spherical symmetry, i.e. the CMB distribution has an intrinsic non-Dopplerian ingredient. This possibility is in a sharp contrast with conventional cosmology. The anticipated magnitude of the non-Dopplerian ingredient is well above the level that already has been reliably detected with the available technology.

The difficulty in detecting the anisotropy in the dipole component of the CMB is not in the accuracy of measurements, but in separation from Doppler's effect. For a monochromatic radiation in the absence of a reference frequency to draw a distinction between Dopplerian and non-Dopplerian shifts in frequency is impossible. For the black body spectrum changes in the temperature parameter caused by Doppler's effect can be incurred by some aperture manipulations. The non-Dopplerian ingredient under aperture manipulations remains intact. As to the Dopplerian component, the performed mathematical analysis comparing aperture effects in straight and reverse directions shows that the resulting temperature shift can be quite significant.

The suspected anisotropy of the CMB can be indirectly deduced from observations of annual variations in the CMB temperature caused by the rotation of the Earth around the Sun. This anisotropy might have been already noticed in what has been considered as non-kinematical part of the CMB dipole but did not interpreted in a proper way.

Direct testing the hypothesis that CMB dipole contains a substantial non-Dopplerian ingredient presents a compelling *experimentum crucis*. The worldview of modern cosmology would hardly sustain if the test results come positive. On the other hand, the suggested model of the physical Universe necessitates the dipole anisotropy of the CMB and readily embodies this fact.

## 1.1. An unexpected prediction with a biological connection

The alleged non-Dopplerian ingredient in the CMB may have an unexpected connection to the mechanism of human perception. For an unprepared reader the idea of such connection looks preposterous.

Isolating the pure Dopplerian ingredient in the CMB dipole would allow determining the true absolute motion of the solar system. The suggested model makes the following prediction: if the non-Dopplerian ingredient in the CMB dipole is subtracted from the vector of the whole CMB dipole, then the remaining dipole vector of the genuine Doppler ingredient will aim in the direction of Virgo cluster.

The rationale for this prediction is based on a minuscule mostly neglected fact: among the whole totality of redshifted galaxies in the Universe there are a few blueshifted galaxies "in a circle of $6^{o}$ radius centered at Virgo cluster" (Peebles, P.J.E., "Principles of Physical Cosmology", Princeton University Press, Princeton, New Jersey, 1993). This means that a tiny fraction of "maverick" galaxies is moving towards the solar system against the overwhelming majority of galaxies that is moving away. This minor exclusion is attributed to an *ad hoc* gravitational force causing an extra peculiar motion due to a hypothetical Great Attractor. The possibility of a non-Dopplerian ingredient in the CMB dipole challenges the entire idea of peculiar motion.

As discussed below, the material world of the Universe develops in a succession of Big Bangs. So, the explanation of the origin of blueshifted galaxies is natural: galaxies from the preceding Big Bang can appear blueshifted since their observed radial velocities can become negative. Thus, if our galaxy is moving with a speed $V_0$ towards a preceding Big Bang galaxy having a lesser speed, $V_p$, then this galaxy will appear blueshifted. The speed of our galaxy is at the lower end of the galaxy speed distribution, so $V_p \approx V_0$. Therefore, the blueshift effect is small in magnitude and numbers. The angular radius of the spot of blueshifted galaxies must be small as well, about *arccos($V_p /V_0$)*.

Thus, the obtained vector of the absolute velocity of the solar system should point towards Virgo Cluster. The vector pointing in this direction is close to the plane of ecliptics and corresponds to the position of autumn equinox in the month of September. It should be noted that in the opposite position is placed the constellation Orion.

The outcomes of the extracorporeal organization of biological information processing may interfere with relocation of a living system in space producing in an effect analogous to astronomical aberration in telescope observations. As described in Appendix B, this "aberration" is supposed to create the famous effect of Moon Illusion – an enlargement on the horizon of the apparent size of the Moon and other celestial objects. Moon Illusion is a psychological not an optical refraction effect as might be convenient to believe. And Moon Illusion turns out to be affected by astronomical conditions of alignment with the absolute velocity of the Earth, which is determined by its rotation around the Sun and the drift towards Virgo cluster with the solar system. Notably, in the most impressive form, the so-called Harvest Moon, this illusion can appear in the middle of September (about a week before the autumn equinox). For other celestial objects, the horizon enlargement is most pronounced for the constellation Orion.

## 2. The Universe as an information processing machine

The properties of the informational infrastructure of the material world have to be in harmony with the great amount of already recognized knowledge about the physical Universe. As said by Richard Feynman: "The problem of creating something which is new, but which is consistent with everything which has been seen before, is one of extreme difficulty." It might be even said that this is practically impossible as long as one sticks to a bottom-up approach - gradual combining of the existing concepts. Great things do not come from small adjustments. The success comes from a top-down design that can reveal a hidden powerful operational principle.

### 2.1. The cellular automaton rule of mutual synchronization

In the middle of nineteenth century with establishing of the wave theory of light had come belief in the notion of a universal ether as a "primordial medium", which was assumed to be the ultimate seat of all physical phenomena. The attempts to understand Nature in terms of ether mechanics were associated with expanding the scope of its activities beyond ponderable matter.

The ether was proposed as the 'true vehicle of life and mind', which can be adapted to a 'spiritual' view of nature (see, e.g. Powers, J., Philosophy and the new physics, Methuen & Co, London and New York, 1985).

Models of ether get fervent support from because they can immediately catch one or another side of physical reality. Thus, considering atoms as vortices in the universal ether J.J. Thomson said: "With reference to the vortex-atom theory, I do not know of any phenomenon which is manifestly incapable of being explained by it." However, numerous ether constructions have failed to develop a comprehensive picture of the physical world. Lord Kelvin said: "We may expect that the time will come when we shall understand the nature of an atom. With great regret I abandon the idea that a mere configuration of motion suffices". In the first place, an ether model of the Universe must address the problem of the organization of the informational infrastructure. Nowadays, with the triumph of information technology the time has come for the idea of information processing facilities of the ether to be carefully evaluated.

There is an inspirational belief that the ultimate construction of Nature should appear in the form of cellular automaton ether. Cellular automata produce sophisticated behavior using simple transformation rules (the most famous example is Conway's "Game of Life"). Still, continual efforts searching for a cellular automaton transformation rule of the physical Universe did not succeed. It is unlikely that such a rule can be discovered by massive trial-and-error efforts in the bottom-up approach. On the other hand, the top-down approach must reveal an operational principle of truly fundamental significance.

Behind any successfully operating system there must be a clock pulse generator. Thus, the primary principle for system design should be: "Cherchez la clock". Actually, the idea of a hidden rhythm that permeates the Universe dates back to Aristotle.

The suggested model of the Universe presents Cellular Automaton EthER InfraStructure (CAETERIS) as a 3-D grid of circular counters. The transformation rule in this cellular automaton model is mutual synchronization in a fault-tolerant mode. This synchronization process creates localized clock pulses that enable operations of the system in a distributed fashion.

Actual realization of a cellular automaton mechanism without driving clock pulses is not possible. In the implementation of cellular automata by computer simulations the indispensable clock pulses do not appear explicitly - the driving clock pulses of programming models are hidden in the clock generator of the computer itself. So, the imperative requirement for a clock generator has been over-looked in search for a cellular automaton rule of the physical Universe by computer simulations.

The suggested CAETERIS model of the Universe utilizes clock pulses from a protocol of mutual fault-tolerant synchronization. The dynamics of this process is described by a parabolic equation for the phase of circular counters with a restriction from below on its spatial derivative. This cellular automaton model produces a set of traveling wave solutions that can be identified with the whole spectrum of stable elementary particles of matter: electron, proton, neutron, photon, and the family of neutrinos. All the properties of the physical world are naturally interpreted in terms of the characteristics of the presented design.

As a point of awareness, it should be mentioned that most of the rule ether models have problems with the polarization of light. In the CAETERIS polarization presents an intrinsic property of the photon synchro-formations.

An elaborated physical theory using partial differential equations is in a sense a study of a cellular automaton model. Thus, hydrodynamics is a study of a model presented by Navier-Stokes equation, electrodynamics is a study of a model presented by Maxwell's equations, quantum mechanics is a study of a model presented by Schrödinger's equation etc. Thus, fundamental theoretical physics as a whole presents a study of a cellular automaton model with a transformation rule of mutual fault-tolerant synchronization. Condensed in one plain sentence the characterization of the Universe goes as follows: "All the phenomena of Nature are various manifestations of activities of mutual synchronization in a network of digital clocks".

Synchronization in distributed systems is a process that is far from trivial and is very rich in consequences. The state of the physical world as a whole is described by a set of phase values in all cellular automaton counters. However, the activities in this system are determined by the differences of these phases by mod $2\pi$. This situation explains the inspirational role in the theoretical physics research of the most effectual symmetry of local and global gauge invariance. And it turns out that structural and behavioral characteristics of the synchronization processes correspond to the traits of the physical world. Especially impressive is the inherent property of slight asymmetry between matter and antimatter with the irreversibility of time at the deepest level of Nature.

An extensive analysis of the cellular automaton mechanism of mutual synchronization has been undertaken. However, delving into minute particulars of the behavior of synchro formations as elementary constituents of matter is of limited interest for biological systems. Biological systems are primarily concerned with the interface to the informational infrastructure of the physical Universe.

## 2.2. Entry to the informational infrastructure of the material world

The phenomena of Nature are divided into three major categories - "Small", "Large", and "Complicated". The transition from the "Small" to the "Large" contains a conceptual gap. Of course, something "Intermediate" should be expected to fill in this gap. But why and how gets it so "Complicated"?

The answer to the "why" part of the question lies in the involvement of the information processing resources below the cellular automaton ether of the physical world. Every biological object gets an access to a "virtual" computer device with a stored program. The answer to the "how" part of the question is related to the organization of biological information processing.

The kinematical scheme of the CAETERIS model is similar to that of the alternative cosmology by E.A. Milne (Milne, E.A., "Relativity, Gravitation and World Structure", Oxford, Clarendon Press, 1935). The distinction of the CAETERIS model is that it includes facilities to store, retrieve, and accumulate a tremendous amount of information required for the realization of Life.

The organization of living systems needs a fast global mechanisms for (1) reference waves for the holographic memory extending through the whole Universe and (2) driving clocks inside each individual biological object. Conventional physics does not have these kind of mechanisms. The CAETERIS model offers an implementation of these mechanisms through a process of "action-at-the-distance".

The concept of the "action-at-the-distance" appears as a formal mathematical trick in Newton's formulation of gravitation and in the theory of quantum mechanics. From a commonsense standpoint, material objects can impact each other only by "action-by-contact", i.e. through a particle or a field. A. Einstein had rejected of quantum mechanics since long range correlations in what is called Einstein-Podolsky-Rosen (EPR) experiment are incompatible with the speed of light limitation. However, experiments show that the EPR paradox indeed take place. Also, "action-at-the-distance" comes out in the Pauli exclusion principle, which is in the heart of the formation of all material substances. Nevertheless, it is not clear how "action-at-the-distance" can be accommodated in contemporary physics and scientists are afraid to consider this concept seriously.

Anyhow, a complete model of the physical Universe must incorporate non-locality. R. Feynman opposed the idea of cellular automaton ether because the locality of the transformation rule could not be reconciled with the non-locality of quantum mechanics. Luckily, the CAETERIS model clearly exposes the origin of the "action-at-the-distance" in the Universe.

The cellular automaton mechanism of mutual synchronization produces two different types of solutions: traveling waves and fast propagating diffusional processes. The former are associated with material formations, the latter arise as a feeble "action-at-the-distance". Parabolic equations have a curious property - the spread of diffusion occurs with an infinite speed. This means that the description of the diffusion mechanism by parabolic equations is a mathematical idealization, which in a strict physical sense is insufficient. In fact, there should be some fast processes behind the simplified representation of diffusion by a parabolic equation. The fast spreading of diffusional solutions in CAETERIS could occur with a speed that is about $10^{40}$ greater than the speed of light. According to our estimates, it would take a diffusional impact about $10^{-22}$ sec to propagate through the whole Universe. In the time-scale of the material processes this diffusional impact can be considered as instantaneous.

There is nothing outlandish to recognize two classes of activities developing in substantially different time-scales. In the realm of material processes, the diffusional "action-at-the-distance" is associated with gravitation and quantum effects. The phenomenon of Life in the physical Universe becomes possible as the diffusional "action-at-the-distance" provides the interface between the material world and the informational infrastructure. In this context, it is interesting to contemplate a remark by A. Eddington that "gravitation propagates with the speed of thought".

## 2.3. The global geometry

The interconnecting links of the CAETERIS elements constitute a 3D topological structure. The emerging material synchro formations in the form of helicoidal traveling waves bring up standards of length and duration along with an upper bound on their propagation speed. Thus, in a local scale the material world is described in the framework of 3D Euclidean metric space with relativistic restrictions on the timing of the events.

The largest part of physics is not affected by the global geometry of the Universe. "Local knowledge cannot give knowledge of the Universe" (E. Borel). However, biological information processing deals with the infrastructure of the physical world as a whole. Therefore, the global geometry of the CAETERIS is determined by operational requirements of the physical Universe as an information processing machine.

Holographic storage of the Universe has to be finite and unbounded. "Holography does not like boundaries" (K. Pribram). The wave processes are most effective when the holographic medium is three-dimensional. This necessitates a conclusion that the CAETERIS is shaped globally as a 3D-hypersphere - a 3D surface of a 4D sphere. The idea that the Universe has such a perfect structure raised fascination through all the times dating back to antiquity: "The nature of God is a circle of which the center is everywhere and the circumference is nowhere" (Empedocles). As we do not have a direct intuition of a four-dimensional space the global structure of the Universe as a 3D-hypersphere can be best understood in comparison with the Earth. The structure of the Earth is a 2D surface of a 3D sphere: locally it appears as a 2D plane although globally it is finite and borderless.

The dipole anisotropy of the CMB results from an eccentric position of observation. Analysis of this configuration allows estimating the parameters of this global structure of the Universe. According to these estimates, the radius of the Universe - the radius of the 4D sphere surrounded by the 3D hypersurface - is about 17.5 billions light years.

## 2.4. Reference waves of the holographic mechanism

As said in (Weyl, H., "Philosophy of Mathematics and Natural Science", Princeton University Press, 1949): "The construction of the world seems to be based on two pure numbers, $\alpha$ and $\varepsilon$, whose mystery we have not yet penetrated." The factor $\alpha = 1/137$, the fine structure constant, appears in relation to interaction of matter with electromagnetic radiation. The number $\varepsilon \sim 10^{40}$ characterizing the relative strength of gravitational interaction is more mysterious: "A simple mathematical theory may lead to numbers like ½ or $8\pi$, but hardly to a non-dimensional number of extravagant order of magnitude $10^{41}$".

Remarkably, the CAETERIS model can explain the mystery of the $\varepsilon$ number. The $\alpha$ number characterizes the traveling wave solutions of the cellular automaton mechanism itself; the $\varepsilon$ number characterizes the diffusional activities in its infrastructure. Both of the numbers, $\alpha$ and $\varepsilon$, present artifacts in the design of the system of the physical Universe: $\alpha$ is determined by the details of the synchronization protocol, $\varepsilon$ is determined by the deep machinery of spreading the diffusional impact. The number $\varepsilon$ characterizes the relative speed of the diffusional impact, which is treated as instantaneous "action-at-the-distance".

Tessellation of the spherical hypersurface of the Universe by the grid of cellular automaton elements specifies two opposite points analogous to the poles on the Earth. The poles of this 3D-hypersurface are the promoting points of global activities in the Universe with periods relating as the pure numbers $\alpha$ and $\varepsilon$. The number $\alpha$ is associated with the Big Bang activities. The number $\varepsilon$ is associated with the information processing facilities of the Universe

The implementation of the reference holographic waves is based on the interplay of global processes of synchronization and desynchronization.

In this connection, it should be mentioned about the problem with the so-called "dark matter". This idea arose first from observations of a "missing gravitational attraction" in some galaxy configurations, then this effect was attributed to a hypothetical "dark matter" for the sake of adjusting the paradigm of the expanding spacetime. The CAETERIS model produces additional sources of gravitational attraction as a secondary effect. This secondary effect is a result of interaction of spreading diffusional solutions from different material objects followed by desynchronization. Such an effect appears only in a large scale forming secondary sources of gravitation in specific halo-shaped areas. So, the CAETERIS model creates some extra gravitational tug, and such a thing as "dark matter" does not exist at all.

The analysis of "dark matter" configurations brought a conclusion that the desynchronization process develops at least about $10^8$ slower than the preceding synchronization process of the spreading diffusional impact. Thus, in the scale of the whole Universe the global processes of synchronization-desynchronization can be repeated with a period of about $10^{-11}$ sec.

The frequency of $\sim 10^{11}$ Hz plays a very special role in the organization of the Universe. At this frequency the global process of synchronization-desynchronization permeates the whole Universe. The biological function of this process is twofold. First, it creates reference waves for the holographic mechanism of the physical Universe that constitutes the basis for memory operations in biological information processing. Second, the passage of synchronization-desynchronization processes through every living being provides an internal clock with driving pulses having $10^{11}$ repetition rate. The internal rhythm of clock pulses in biological objects at $10^{11}$ Hz can be influenced by external electromagnetic radiation with close frequency. Electromagnetic radiation at this frequency produces a variety of strange biological effects, although they are weak and apparently harmless.

In the physical world, the frequency of about $10^{11}$ Hz presents a watershed between quantum and classical effects. In a global scale, the synchronization-desynchronization processes undulate the shock wave accompanied the Big Bang resulting in the blackbody spectrum of CMB with 2.72 $^o$K temperature. According to Wien's displacement law, the maximal intensity of the black body spectrum with a parameter T= 2.72 $^o$K is attained at the frequency of about $10^{11}$ Hz.

## 3. Subsystems of the Universe and global periodic processes

In brief, the CAETERIS model of the Universe presents a network of circular counters interconnected locally by a 3D pattern. The dynamics of the network is determined by a cellular automaton rule of mutual synchronization in a fault-tolerant mode. The general construction of the network is in the form of a hypersurface – a 3D surface of a 4D sphere. Two opposite poles of this hypersurface carry an essential operational function in generating periodic activities in the global scale.

The two pure numbers, $\alpha = 1/137$ and $\varepsilon \sim 10^{40}$, correspond to typical speeds of different activities of the physical world. In association with the pure numbers $\alpha$, 1, and $\varepsilon$, the global processes in the Universe fall into three categories of subsystems. These subsystems have different compositions, exhibit different types of behavior, and develop in different time scales.

## 3.1. The subsystem of material formations

The spectrum of travelling wave solutions of the CAETERIS model yields synchro-formations that can be identified with all the stationary elementary particles: electron, proton, neutron, photon, and a family of neutrinos exhibiting corresponding properties of mass, charge, and spin. Being in a helicoidal form, these synchro-formations can be presented in by means of dual solutions, which are related to antimatter. Slight asymmetry in the foundation of Nature, the intrinsic irreversibility of the physical time and the CP violation, are determined by the requirement of arbitration of switching signals in the cellular automaton elements. The dynamics of material synchro-formations exhibits three fundamental interactions. Gravitation is a separate effect related to fast propagating diffusional solutions.

The travelling wave synchro-formations exist only in motion being propelled by the underlying cellular automaton mechanism. Thus, the property of inertia is intrinsic to material formations as they get the ability of uniform motion for "free".

An upper bound on the propagation speed of synchro-formations is determined by a fixed minimal value in phase change for the mutually synchronizing circular counters of the cellular automaton elements. This requirement ensures a fault-tolerant mode of operation of the cellular automaton mechanism. The minimal phase change establishes metric standards in space and time as a measure of a full $2\pi$ phase reverse in extent of nodes for the measure of distance and in amount of counting cycles for the duration of time. The ratio of these two intrinsic numbers of the system presents $c$ - the upper bound on speed of the travelling waves synchro-formations.

On the other hand, there is a lower bound for the speed of the traveling wave solutions in connection to the stability of the created material synchro-formations. This lower bound, $V_{matter}$, is a fraction of $c$ determined by the fine structure constant $\alpha$: $V_{matter} = c \cdot \alpha = c \cdot 1/137$. The value of $V_{matter}$ is the speed of an electron at the first orbit of Bohr's atom.

## 3.2. The subsystem of electromagnetic radiation

The fast moving traveling wave solutions at the speed $c$ are simpler and have less diversified structures than slower material formations. These fast moving structures include material synchro-formations for photons and neutrinos, and electromagnetic radiation. The processes at the speed $c$ serve for transmitting interactions of slower material formations. When slow moving material formations are transformed into fast formations at the speed $c$ the disappearing mass, m, of the slow formations converts into a thrust of the fast formations in proportion $\sim mc^2$.

The structure of the electromagnetic waves as portrayed by the scheme of Maxwell's equations exists up to the frequency of about $10^{11}$ Hz. Above this frequency the electromagnetic radiation is replaced with traveling wave solutions of photons. Photons move with the same speed $c$ but have a structure completely different from that of electromagnetic waves. The frequency of about $10^{11}$ Hz presents a watershed between quantum and classical effects.

## 3.3. The subsystem of feeble impacts of the "action-at-the-distance"

Diffusional solutions of extremely rapid "action-at-the-distance" pervades the material world of "slow" and "fast" travelling synchro-formations. At the macro-level, the diffusional solutions manifest the omnipresent attraction of gravity. At the micro-level, the diffusional solutions being modulated by material formations serve as short-term storage of information. In this way, they set up information preprocessing that governs the dynamics of microobjects. Having very fast information preprocessing as an initial step in quantum transitions is a decisive determinant of quantum mechanics behavior.

In the global scale, diffusional solutions from different aggregations of material formations can interact to give rise to secondary sources gravitation. These secondary sources appear from an interplay between synchronization and desynchronization processes. The spreading diffusional solutions determined by material synchro-formations intersect in a typical shape of a halo considered as a location of "dark matter".

For the organization of living systems this subsystem plays a vital role in providing reference waves for the holographic mechanism of biological information processing.

### 3.4. Two large-scale periodic processes

Global activities in the Universe arise as periodic processes that can develop in a self-oscillation mode between the poles of the cellular automaton structure. Periodic process as an object of scientific study has an epistemological advantage over a unique singular event.

### (1) Creation of matter

In the CAETERIS model elementary constituents of matter present relocating patterns of synchro formations. Regularly, they are not created or destroyed, they are just transforming from one configuration to another. The regime of conservation of matter is escaped when the cellular automaton procedure of mutual synchronization is breached. Abrupt forced change of the phase at the pole results in outburst of helicoidal kernels of travelling wave solutions. These synchro formations are elementary constituents of matter, or antimatter, depending on the sense of rotation. The choice of the sense of rotation is wired in the cellular automaton an arbitration protocol to resolve the order for processing of simultaneously arriving signals. Two different situations are indistinguishable: (1) signals are processed according to the order of moments of their arrival or (2) simultaneously arriving signals are processed according to the arbitration rule. Thus, the irreversibility of time is intrinsically involved in matter-antimatter creation. This elucidates one of the cardinal problems of fundamental physics – a slight asymmetry of the physical world related to the so-called CPT invariance.

In the CAETERIS model, the CMB radiation appears as a factor that accompanies the outburst of material synchro formations. Initiation of material synchro formations starts with abrupt change of the phase at the pole point. Besides creating material synchro formations changing the phase at the pole point is accompanied with a "shock wave". This "shock wave" has a spherical shape and embraces the "fireball" of the seeds of matter.

### (2) Undulation of synchronization – desynchronization

The speed of the diffusional impact is determined to be $V_{diffusion} = c \cdot 10^{40}$. It takes this impact about $10^{-22}$ sec to travel across the whole Universe of radius 17.5 billions light years. Thus, the synchronization impact spreading from one pole of the Universe reaches the other pole in a very short time. This synchronization state will desynchronize and reach the opposite pole at a relatively slower pace. Then, the cycle of synchronization and desynchronization repeats. Since the synchronization impact is very rapid, the repetition rate of undulations is determined by the speed of desynchronization and is supposed to be about $10^{11}$ Hz.

The speed of the desynchronization process can be evaluated in conjunction with the analysis of secondary sources of gravitation considered as "dark matter".

## 4. An alternative interpretation of cosmological processes

Despite its primary orientation at micro-physics and organization of biological information processing, the CAETERIS model has been effectively elaborated to incorporate fundamental notions of cosmology. It turns out that this model can provide a natural harmonious explanation to a number of apparently unrelated controversial astrophysical observations.

### 4.1. Recurrent creation of matter in a periodic sequence of Big Bangs

Creation of material synchro formations goes together with a spherical "shock wave" of the Cosmic Microwave Background (CMB). Travelling with the speed of light, this "shock wave" gets from one pole to another in about 60 billion years (presuming the estimated radius of the Universe of about 17.5 billions light years). The "shock wave" reaches the opposite pole in a convergent manner forcing a fixation of the phase at this point. This provides a condition for another Big Bang activity of creation material synchro formations. The arbitration protocols at opposite poles may mirror each other, so while one pole produces matter the other produces antimatter.

This scenario suggests that creation of material formations in the Universe is not a unique event of an isolated Big Bang. Instead, the creation of matter in the Universe is a periodic process occurring in a sequence of Big Bangs. The Big Bangs alternate between creation matter and antimatter at intervals of about 60 billion years. The pole at our side of the Universe is set up to generate Big Bangs of matter, the opposite pole creates antimatter. Thus, we may have several Big Bangs of matter in front of our Big Bang. At the "equator" area of the Universe the matter and antimatter Big Bangs coming from different extremes may collide.

The presented scenario brings to resolution two momentous astrophysical conundrums. Also, it instigates an intriguing biological remark.

### (1) What is going on in the faraway cosmos?

In conventional cosmology, the early Universe should present a quiet zone of transition from "smooth CMB" to "lumpy" stellar systems. Unexpectedly, this is not the case: the faraway cosmos is full of the most energetic activities of the whole Universe. In the presented scenario, the faraway cosmos gets a continuous supply of energy from annihilation of colliding matter and antimatter. Sporadic gamma-ray bursts occur from collisions of individual stars, lasting quasars occur from collisions of galaxies. High redshift quasars are frequently aligned with low redshift galaxies. This fact looks very paradoxical as such redshift differences imply billions of light years separation (see Arp, H., "Quasars, Redshifts, and Controversies", Interstellar Media, Berkley, California, 1987). Yet this alignment can occur in collisions of larger galaxies of matter with smaller galaxies of antimatter, so the created quasars would acquire a higher redshift of kinematical origin.

**(2) Do the galaxies rush apart in an accelerated manner?**

Recent astrophysical observations have revealed that certain stellar formations have smaller velocities than their distances prescribe. This has been interpreted that galaxies are pulled away not with steady velocities, but in an accelerated manner. Therefore, to keep up with the paradigm of the expanding spacetime of general relativity it becomes necessary to assume that some unknown kind of "dark energy" dominates the Universe.

However, the results of these confounding observations are in superb agreement with the suggested model of the physical Universe. In a model of the Universe presenting a periodic sequence of Big Bangs galaxy positions in the physical space and the redshift space may be reversed. Thus, the far-away slow objects are galaxies that had been materialized in a previous Big Bang, fast objects that are nearby merely belong to the current Big Bang.

**(3) A biological remark**

The idea that the creation of matter in the Universe is a recurrent process rather than a singular event is of great significance for biology. The conventional cosmological paradigm makes available only a relatively short period of time of about 4 billions years for the development of Life on Earth from the inert matter. In the CAETERIS model, the assets of Life are in the informational contents of the Universe infrastructure. So, the development of living systems and accumulation of biological information in the Universe may had started many hundreds billion years before the latest Big Bang produced the current instance of the material world.

**4.2. Why the CMB has a spectrum of blackbody radiation at 2.72 $K^0$ ?**

In the CAETERIS model the Cosmic Background Radiation (CMB) is not a post-creation remnant of cooling down matter but an accompanying factor of the Big Bang resulting from the "shock wave". In conventional cosmology the 2.72 $K^0$ temperature of the CMB is an arbitrary parameter depending on the epoch of observation. In the CAETERIS model this parameter is of completely different significance. The 2.72 $K^0$ temperature of the CMB is a design parameter of CAETERIS, which is determined by the frequency of the reference wave of the holographic mechanism of the Universe.

As has been indicated, the maximum intensity in the black body spectrum with the temperature parameter 2.72 $K^0$ corresponds to the frequency $10^{11}$ Hz. The spherical "shock wave" that embraces the Big Bang aggregation of matter is affected by the synchronization-desynchronization undulations at about $10^{11}$ Hz that permeates through the whole Universe. One can imagine that alterations of synchronization-desynchronization can scatter the shock wave vibrations and create photons. Following the Plank's scheme of derivation of the spectrum of the black body radiation it is possible to have a similar scheme for statistical distribution of the photons from the shock wave vibrations. The CMB does not present the black body radiation, merely the CMB spectrum originates from the same statistical distribution scheme.

**4.3. An eccentric view on receding galaxies**

Creation of matter in a single point explosion corresponds to Milne's cosmological model. At the first sight, it seems that in a single point explosion the Hubble picture of receding galaxies relates exclusively to the center in contradiction to the cosmological principle. However, the core issue of Milne's model is that the same Hubble picture of flying apart galaxies is seen not only from the center but from any of the receding galaxies. For some reasons, this paradoxical circumstance is not widely recognized.

Thus, it is important that an eccentric position of observation does not distort the Hubble picture of receding galaxies. But, an eccentric position of observation distorts the spherical symmetry of the temperature distribution of the CMB. As a result, being observed from our solar system, the spherically symmetric distribution of the CMB temperature acquires a significant non-kinematic dipole component in addition to the dipole component due to Doppler's effect. In conventional cosmology, the whole dipole component of the CMB is attributed to Doppler's effect. This leads to a wrong conclusion about the global motion of galaxy systems in the Universe considering that the motion of the solar system deviates from the Hubble expansion by a powerful tug from a hypothetical Great Attractor. Testing for a non-Dopplerian ingredient in the CMB is a decisive challenge for the notion of the Great Attractor.

In Milne's model material structures are developing in a centralized way while in conventional cosmology material structures precipitate from the expanding spacetime in a decentralized way. Practicing the theory of the Universe with a decentralized creation of material structures requires an *ad hoc* artificially invented stage of initial inflation.

Furthermore, in decentralized creation spatial distribution of galaxies is expected to be uniform. However, observations reveal large coherent structures in the nearby Universe: the distribution of galaxies is not only non-uniform but non-random. The spatial distribution of galaxies exhibits well pronounced patterns such as: extended sheets (the "Great Wall"), immense spherical and tube-like voids (not random deficiencies of matter), filaments and cuts through two-dimensional sheets. Such patterns can naturally appear in a centralized model, but in a decentralized environment of the expanding spacetime the emergence of non-random patterns cannot get a consistent explanation. Thus, the problem of appearance of specific configurations in the distribution of galaxies is basically ignored.

## 5. Quantum mechanics and biological information processing

It has been anticipated for a long time that the mystery of quantum mechanics and the mystery of Life are interrelated. Both quantum and biological phenomena are associated with the informational infrastructure of the physical world using the "action-at-the-distance" mechanism of the CAETERIS construction. Biological control appears just as a more sophisticated version of the regulation of quantum transitions. The substantial amount of this sophistication establishes the distinction between dead and living matter.

First of all, quantum and biological phenomena lean on information processing facilities of the Universe in different time scales with respect to the critical frequency $10^{11}$ Hz. Basically, quantum processes develop in the time scale below $10^{-11}$ sec. Thus, quantum states of macroscopic quantum effects can be destroyed by impacts of electromagnetic radiation at about $10^{11}$ Hz. Biological processes are driven in the time scale above $10^{-11}$ sec as long as the critical frequency $10^{11}$ Hz determines the driving clock inside all biological objects.

The different time scales of quantum and biological processes affect the involvement of the memory mechanisms of the Universe. Being associated with frequencies above the critical frequency $10^{11}$ Hz of the reference waves, quantum processes do not utilize the holographic storage of Universe's infrastructure. Following Feynman's formalism of trajectory integrals, the organization of quantum transitions relies on the short-term memory property of phase conservatism of mutual synchronization. On the other side, biological processes developing below the critical frequency $10^{11}$ Hz can exploit the facilities of the holographic memory.

Quantum mechanics as a theory of micro-objects is very precise, but exhibits a number of peculiar features in their behavior. These features fall into three groupings: wave-particle duality (a micro-object appears as a wave or as a particle depending on the experimental setup), "non – objectivity" of measurements (micro-objects move without a trajectory and can be detected in arbitrary place), and long–distance correlations (entangled quantum objects can affect each other in far away locations). These features seem very strange and appear unrelated.

Ongoing attempts to penetrate the mystery of quantum mechanics go around a supposition that any trait in the behavior of material objects must result from one or another type of immediate physical interaction. The CAETERIS model brings up a different approach to this problem attracting information processing. Quantum transitions are presented as a two–step process: almost instantaneous information preprocessing and a relatively slow material development resulting in an actual material transition. The step #1 lays down the groundwork in a wave-like information preprocessing for a particle-like motion at the step #2. The outcomes of quantum measurements are not inherent to the micro-objects per se but are determined by the way how the information scene for the transition is set up. And since information preprocessing is done instantaneously, measurements for a system of several micro-objects would reveal correlated parameters irrespectively of how far these micro-objects are separated.

So, the dynamics of micro-objects develops as follows: a preliminary very fast preprocessing of information paves the subsequent actualization of material events. A two-step scheme with a fast preprocessing stage perfectly elucidates all the strange aspects of quantum mechanics behavior and makes them crystal-clear.

Analogously, a two-step arrangement in biological information processing also relies on the same "action-at-the-distance" mechanism for fast preliminary operation. This scheme provides an effective integrated implementation of human perception.

## 6. Conclusion remarks

The idea that the physical Universe has an immaterial spiritual infrastructure is a popular subject of metaphysical deliberations. The specificity of the given work is that it presents this idea in terms of a concrete engineering design in the framework of information technology.

The approach of engineering design has a methodological advantage over the traditional application of mathematical modeling. Engineering design must take care of the finest details of the developed construction, which in mathematical modeling are omitted.

The design of the Universe presents a Cellular Automaton EthER InfraStructure (CAETERIS) whose global geometry is a 3D hypersurface of a 4D sphere. The Universe undergoes two types of global periodical processes between the poles of this hypersphere: explosive creations of material formations with a period of about 60 billions years and undulations of synchronization-desynchronization with a period of about $10^{-11}$ sec.

The former process presents a succession of Big Bangs supplying the Universe with material formations. The latter process plays a decisive role in the organization of information processing responsible for the emergence of living matter in the physical Universe – it produces reference waves for the holographic mechanism of the Universe and clocks of driving pulses inside every biological object.

There is nothing wrong or "anti-scientific" in considering information dominant Universe. Anyhow, remodeling of modern cosmology is impending because of its own internal conflicts. Apart from cosmology, the significance of general relativity for the rest of physics is minor. Reduced to its essence, the idea of relativity promotes a postulate that any process in the physical Universe is determined only by configurations of the material structures involved. However, the "barcode" interpretation of the DNA functionality clearly indicates that material configurations *per se* do not provide sufficient variety for the development of biological objects.

A. Einstein once said: "I want to know how God created this world. I want to know His thoughts, the rest are details." The design of the Universe has to provide operational support to the major phenomenon of this world  - the phenomenon of Life. From the standpoint of this design many questions that look profound for theoretical physics are details of minor significance. Perhaps, such details are valuable along the way of a bottom-up construction. But as a general model is established through a top-down design delving into minute particulars beyond the point of model validation may be worthless.

The Universe as an information processing machine is characterized in Fig. A.

## 7. Annotated bibliography

1. The idea of modeling physical Universe using cellular automata has been ardently advocated by E. Fredkin; it has been investigated  by T. Tofoli, N. Margolis, G. Vichniac, S. Wolfram, and others (see, e.g.,  S. Levy, "Artificial Life", Vintage Books, New York, 1992).

2. The cellular automaton model of the physical Universe based on a mechanism of mutual synchronization has been described in publications by S. Berkovich:

   - "Mutual Synchronization in a Network of  Digital Clocks  as the Key Cellular
      Automaton Mechanism of Nature.  Computational Model of Fundamental Physics",
      Synopsis,  Rockville, MD, 1986,   ISBN   0-9613945-1-X
   - "Cellular automaton modeling of the phenomena of fundamental physics",
      Proceedings of the 19[th] Annual Pittsburgh Conference on Modeling and Simulation,
      vol. 19, Part 2, pp. 895-906, 1988
   - "Spacetime and matter in cellular automaton framework",
      Nuclear Physics B (Proc. Suppl.) 6 (1989) 452-454
   - "A possible explanation of quantum mechanics behavior by a classical cellular
      automaton construction", Bell's theorem, Quantum Theory and Conceptions of the
      Universe (M.Kafatos, ed.), Kluwer Academic Publishers, 1989

3. A distributed fault-tolerant synchronization protocol that entails a specific lower bound restriction on the change of clocks phase has been presented in S. Berkovich, S. Haaser, H. Yee, and C. Walter, "Distributed Multiple Clock System and a Method for the Synchronization of a Distributed Multiple Clock System", US Patent No. 5,295,257, Date – Mar. 15, 1994

4. The analysis of the model, which is dubbed in this paper CAETERIS (Cellular Automaton EThER InfraStructure), has been scattered in numerous publications.
   An overview of the efforts for the development of this model can be obtained from the abstracts in the Bulletin of American Physical Society (BAPS):

   - "Elementary particles of matter as cellular-automaton formations of mutual synchronization", BAPS, Vol. 32, No. 6, p. 1435, 1987
   - "Cellular automaton approach to grand unification", BAPS, Vol. 33, No. 4, p. 1096, 1988
   - "Computational model of fundamental physics", BAPS, Vol. 34, No. 5, p. 1429, 1989
   - "On the interpretation of quarks as time-alternating states", BAPS, Vol. 35, No. 4, p. 951, 1990 (co-author V.Krasnopolsky)
   - "Spontaneous symmetry breaking and anisotropy in CP-violating decays", BAPS, Vol. 36, No. 4, p. 1247, 1991
   - "Is there a cellular automaton ether? – Experimentum crucis", BAPS, Vol. 37, No. 2, p. 966, 1992
   - "Logical Infeasibility of Relative Motion Constructs", BAPS, Vol. 38, No. 2, p. 1004, 1993 (co-author E. Berkovich)
   - "Dark matter as a remote phantom image of massive bodies", BAPS, Vol. 40, No. 2, p. 941, 1995
   - "An eccentric view on the kinematical scheme of the Big Bang", BAPS, Vol. 40, No. 2, p. 942, 1995 (co-author J. Favre)
   - "Concept of relativity and empirical observation of the detachment of mental imagery", BAPS, Vol. 42, No. 2, p. 1129, 1997
   - "Modeling of spatial distribution of galaxies in alternative Milne's cosmology", BAPS, Vol. 42, No. 6, p. 1580, 1997
   - "On the alignment of astronomical objects with disparate redshifts and generation of gamma-ray bursts", BAPS, Vol. 43, No. 2, p. 1143, 1998
   - "On a spot of blueshifted galaxies in the Virgo cluster", BAPS, Vol. 44, No. 1, part II, p. 1405, 1999

5. Publications, reports, and manuscripts related to the analysis of the model have been compiled and translated into Russian by G. Lapir and V. Arshinov of the Institute of Philosophy of Russian Academy of Sciences:

   S.Y. Berkovich, "Cellular Automaton as a Model of Reality: Search for New Representations of Physical and Informational Processes", Moscow University Press, Moscow, Russia, 1993 (in Russian)

   This book contains the most complete description of the physical properties of the CAETERIS model (excluding cosmological topics that have been developed later).

6. For some obscure reasons physical Universe incorporates a very fine effect of a slight

asymmetry between matter and antimatter – the CP violation. This effect is related to the irreversibility of time at the deepest level of Nature. One of the most impressive features of the suggested model based on mutual synchronization principle is that it must incorporate an arbitration mechanism that simultaneously leads to the effects of CP violation and irreversibility of time. The slight asymmetry in the CAETERIS construction is described in: S. Berkovich, "Symmetry Breaking Effects in Cellular Automaton Modeling of the Physical World", Proceedings of the 22$^{nd}$ Annual Pittsburgh Conference on Modeling and Simulation, vol. 22, Part 5, pp. 2523-2532, 1991

7.  One may get the wrong impression that having a cellular automaton infrastructure of the physical world contradicts the fundamental notion of relativity. In the prevalent popular interpretation of relativity, the question of whether there is an absolute frame of reference of the physical world seems to have been resolved once and for all in a negative sense by the Michelson-Morley and similar experiments. However, the concept of relativity can be presented in two interpretations: according to Einstein the absolute frame of reference does not exist, according to Lorentz, Poincaré, and others the absolute frame of reference is simply undetectable. Considering this difference may seem as a scholastic dispute on distinguishing between something that does not exists and something that does exist, but is undetectable. But, strictly speaking, the concept of relativity addresses only the undetectability of uniform translational motion in mechanical, optical, and electromagnetic experiments. The possibility of observing other attributes of absolute space in other types of experiments is not excluded.

   Evidence for a preferred frame of reference has been supposedly provided by observation of the dipole anisotropy in the cosmic background radiation. The apparent contradiction of this fact with the concept of relativity is confusing and requires an explanation as stated, for example, in a note by R. J. Yaes, "Reconciling COBE Data with Relativity", Physics Today, p. 13, March 1993, which says: "measurement of this dipole anisotropy is in effect a modern-day Michelson-Morley experiment, but this time with a positive result, and that the cosmic background radiation acts in effect like the stationary ether that Albert A. Michelson and Edward W. Morley failed to find".

   Another "modern-day Michelson-Morley experiment" for testing the existence of the absolute frame of reference can be devised in relation to the alleged hazardous bio-medical impacts from the electromagnetic fields EMF ( S.Y. Berkovich, "A hypothesis on the origin of health-hazardous EMF effects", in "Project Abstracts, The Annual Review of Research on Biological Effects of Electric and Magnetic Fields." W/L Associates, Savannah, Georgia, pp. 93-94, 1993; also, http://www.seas.gwu.edu/seas/eecs/News/fall94/em.html).
   There is a hypothetical possibility that certain disturbances in cellular automaton activities being left over in an absolute position occupied by one object may influence a trailing object relocating to this absolute position shortly thereafter. These after-effect disturbances do not have to be necessarily small to pass unnoticed in common laboratory practice. Being unexpected, they would be disregarded as strange transient flukes unless an intentional search has been undertaken. Without a special endeavor these after-effects can reveal themselves under two conditions: if their actions are accumulated and if their outcomes are subject to registration. The situation with the hazardous EMF effects represents an exclusive case where both of these conditions are satisfied.

   For about two decades, the investigation of the alleged health hazards from electro-magnetic fields (EMF) (see, e.g. . W.R. Bennett, "Cancer and Power Lines", Physics Today,

April 1994, pp. 23-29) has been caught in a confrontation between physical and epidemiological judgements. On one hand, it appears that being small in a relative and absolute sense magnetic fields from power lines cannot produce a discernible bio-medical effect. On the other hand, numerous studies consistently show a "weak, but statistically significant" link of power lines with some harmful effects like childhood leukemia.

Biological impact from power lines can be due some factor other than EMFs. Actually, epidemiological studies show an association of the observed effects with calculated magnetic fields rather than with measured fields. Trusting in epidemiological and physical analysis, one comes to the conclusion that the carcinogenic action attributed to EMF's is determined by the proximity to electrical wires of which the magnetic field is only an indicator.

Ordinarily, material objects are assumed to influence each other through mediative agents. After-effect hypothesis offer a simple possibility for a non-mediative impact. Assume that a material object creates some disturbances in its position in the infrastructure underlying the physical world and that these disturbances can influence a trailing object which relocates to this position shortly thereafter. In experimental testing, biological objects have to be exposed to the presence of high-voltage lines not to "an equivalent magnetic field".

Thus, the health hazardous effects from electrical wires attributed to EMFs may reveal a novel facet of reality: dependence of some kind of events on the absolute positioning of the material objects involved. Thus, favoring the Lorentz-Poincaré philosophy could open new avenues in biology while leaving Einsteinian physics intact.

8.  The properties of the CAETERIS model in cosmological aspect have been described in three technical reports of the Institute for Information Science and Technology of the George Washington University:

1.  S. Berkovich, "Dark Matter as Remote Phantom Image of Massive Bodies", GWU-IIST 94-06, March 1994

Physical Universe exhibits a puzzling trait of excessive gravitational attraction beyond the scope of visible matter. A straightforward approach would be to postulate that this effect is due to "dark matter" - a sort of customary matter, which is deprived of the property of luminosity. The *ad hoc* postulate of "dark matter" raises a variety of problems of physical origin. Besides that, the postulate of "dark matter" does not account for its specific spatial configurations. In the presented model, the observed extra gravitational attraction appears as a secondary effect resulting from phantom imaging of massive bodies. Primary synchro activities of ponderable matter are associated with helicoidal kernels that create stretching diffusional solutions in cylindrical form (associated with electrostatic field). This process develops as fast front propagation of synchronization followed by a relatively slower process of desynchronization. The secondary sources of gravitation are created at the points of intersection of primary activities. Although the density of primary activities falls inversely proportional to the square of the distance their operative time (difference between passage of synchronization and desynchronization) increases linearly with the distance. Therefore, the secondary effects of gravitation do not appear on short distances and manifest only at sufficiently large distances where the operative time becomes effectively long. On the other hand, the density of these secondary effects diminishes at very large distances. So, they do not appear in intergalactic space. The described secondary effects of gravitation produce typical geometrical patterns of hallo surrounding spiral galaxies.

2. S. Berkovich & J. Favre, "An Eccentric View on the Kinematical Scheme of the Big Bang and the Absolute Impression of the Cosmic Background Radiation", GWU-IIST 96-04, January 1996

The works presents the CAETERIS model in view of the alternative cosmological model of E. Milne. The eccentric position of observation preserves the Hubble picture of the receding galaxies but reveals particularities in temperature distribution of the CMB. The distinctive feature of the CMB in eccentric observation is a non-Dopplerian dipole ingredient. Looking for this non-Dopplerian ingredient constitutes an *experimentum crucis* for the presented concept. The work describes a procedure that may distinguish between dipole ingredients of Dopplerian and non-Dopplerian origin. The analysis suggests a particular placement for the quadrupole component of the CMB.

3. S. Berkovich, "On the Kinematical Scheme of Alternative Milne's Cosmology: Architecture of Galaxy Distribution and Generation of Gamma-Ray Bursts in Annihilation Clashes between Successive Big Bangs", GWU-IIST 97-01

The objective of this work was to develop a comprehensive description of large-scale structures in the Universe. In conventional cosmology the distribution of galaxies must be uniform. But observations reveal characteristic patterns in galaxy distribution such as: extended sheets, immense voids, filaments and cuts through two-dimensional sheets. With the suggested scheme, all these patterns can be reproduced effortlessly in computer simulations. A natural extension of this scheme leads to a hypothesis of successive Big Bangs providing an insight into Deep Hubble Field. The high-energy activities in this area that is supposed to be quiet arise from the clash of the matter and anti-matter Big Bangs. As the slow far-away objects can come from the previous Big Bang, the scheme of successive Big Bangs easily resolves the controversy with "accelerated" expansion of the Universe.

9. Total number of known species of living organisms on Earth is about 1.75 million ("Scientists to create 'Catalogue of Life'", Scientific Computing World, July/August 2001, p. 9). Just to start the discussion let us assume that there are 10 billion living organisms of each specie. Also, let us assume that the cells of each specie contain 50 chromosomes. An estimate of the total number of living beings on Earth is about $2 \cdot 10^{16}$, and the total number of transmitters-receivers as determined by the number of chromosomes is about $10^{18}$. This value is far below $- 10^{33}$ - the evaluated number of users of "the Internet of the physical Universe" (Fig. A). Microorganisms and low-level organisms can share DNA keys.

This estimate shows two basic things: (1) sharing the bandwidth of the underlying informational infrastructure of the physical world for the organization of Life as a collective activity is feasible and (2) a more accurate estimate can bring a conclusion whether a sufficient bandwidth is still unused to warrant Life in other places of the Universe. This can provide some quantitative guidance to the project SETI (Search for Extra-Terrestrial Intelligence).

# PHYSICAL  UNIVERSE  AS  A  CYBERSPACE

| | |
|---|---|
| *INFORMATIONAL  CAPACITY:* | --- $10^{102}$ PHYSICAL BITS |
| | --- $10^{100}$ LOGICAL BITS |
| *TOTAL  NUMBER  OF  LIVING  BEINGS*<br>*( DEAD  AND  ALIVE )* | --- $10^{80}$ |
| *THE  TOTAL  VARIETY  OF  ACCESS  CODES* | --- $10^{30}$ |
| *CLOCK  RATE* | --- $10^{11}$ Hz |
| *WORKING  FREQUENCY  OF  REFERENCE  WAVES*<br>*(COMPARE TO PLANK'S  TIME: $h \times G / c^5 = 5.3 \times 10^{-44}$ SEC)* | --- $10^{46}$ Hz |
| *MODULATION  FREQUENCY* | --- $10^{13}$ Hz |
| *NUMBER  OF  SIMULTANEOUS  USERS* | --- $10^{33}$ |

*HUMAN  BRAIN  AS  A  NETWORK  COMPUTER*

**OPERATIONAL  FEATURES:**

- DISTRIBUTED CONTENT – ADDRESSABLE ACCESS
- MERGING OF DATA AND INSTRUCTIONS
- RANKING SUBSETS FROM MULTIPLE RESPONSES
- NON-ERASABLE ACCRETION OF INFORMATION
- PROLIFERATION OF VARIOUS COPIES OF DATA
- MULTIPROGRAMMING BY INTERLEAVING OF ATOMIC OPERATIONS
- INTEGRATION OF INPUTS THROUGH DIRECT MEMORY ACCESS

**MEMORY  PROPERTIES:**

- SINGLE LEVEL
- STORAGE CAPACITY IS VIRTUALLY UNLIMITED
- FAST READ/WRITE OPERATIONS
- HIGH THROUGHPUT
- GREAT RELIABILITY AND DURABILITY

Fig. A

Characterization of biological information processing

## Appendix B.   Computational model for biological information processing

Real understanding of Life requires an operational explanation rather than a "responsible for" type of description. Such an explanation has to come in the form of a universal algorithmic construction to embrace the whole organization of biological information control.

The situation with biological information processing is confusing in many respects. The problem is not only with the deficiency of appropriate hardware. Assuming that any required hardware is readily available the question about the structure of software still remains unclear.

Software design starts with the identification of a computational model - an abstract scheme for implementation of algorithms, which presents a set of primitive constructs and rules for manipulation of information items. The concept of a computational model is in the heart of algorithmic developments. The Turing Machine is a computational model of pure theoretical interest that formalizes the notion of an algorithm through symbol manipulations in a sequential memory. A practical computational model, introduced by von Neumann, utilizes a random access memory with word-organized operations. Despite tremendous advancements in technology the original von Neumann model stays as the prevailing scheme for modern computer design. According to the so-called Church-Turing Thesis, all reasonable computational models are equivalent in algorithmic capabilities, although they may drastically differ in performance.

Further computational model developments should rely on a different mode of memory access. This leads to the concept of content-addressable storage, which provides a faster access to information items through their contents. Using content-addressable storage gives rise to the problem of resolution of multiple responses. A special mechanism for resolution of multiple responses allows information items to be extracted from the memory in a sequence and processed one at a time. So, a computational model with content-addressable memory having a built-in mechanism for resolution of multiple responses basically operates as an ordinary von Neumann's model, except perhaps for some performance enhancement. This computational model with content-addressable memory got only limited application in computer technology for a number of hardware and software reasons.

A computational model with content-addressable memory that does not include a mechanism of resolution of multiple responses presents quite a different situation. This computational model has to manipulate with subsets rather than with individual data items. Such a situation is typical for holographic content-addressable memory, which does not have a mechanism for resolution of multiple responses. Utilization of a computational model based on content-addressable memory without resolution of multiple responses that has to manipulate with subsets rather than with individual information items is a characteristic distinction of the organization of biological information processing.

The most controversial issue in the proposed organization of biological information processing is that it employs extracorporeal facilities. The "barcode" functionality of the DNA suggests that a biological object can use these facilities by getting access to a slice of holographic storage of the Universe. Thus, before going into consideration of how the new computational model suits the control needs of biological objects it would be sensible to test whether this model indeed relies on extracorporeal organization of information processing.

## 1. *Experimentum Crucis* – Extracorporeal Placing of Human Memory and Visual Perception

The suggested organization of biological information processing has an extremely clear construction feature: the memory involved in this processing is situated outside of the material structures of living systems. This supposition looks utterly absurd from the standpoint of conventional science. Therefore, this circumstance greatly enhances the qualifications of the proposed investigation as an *experimentum crucis*.

Extracorporeal placement of human memory is closely related to the perception of video imagery. Every imprint of a visual image, at least for some time, is retained in memory. There are two possibilities how the visual information gets its way to the memory. Visual perception system can operate as a camcorder that reviews a picture before putting it in a cassette, or visual perception system can operate as a network computer that directs incoming information into the storage and then displays it from the memory. Logical facilities of both of these structures may be very similar, but physical arrangements corresponding to these two schemes can be positively distinguished.

The suggested experimentation deals with afterimages – brain activities following a visual stimulation. Usually, psychology studies do not entail enough precision for conclusions about physical organization of the brain. The presented case is different as it relates to investigations of afterimage effects with closed eyes or in complete darkness. Thus, a head with an afterimage activity inside can be treated as an isolated system apart from concerns about localization of internal processes. Therefore, the obtained results can be subject of a discussion on pure physical grounds.

Although the phenomenon of afterimages is well known its roots may not be properly understood. The meaning of the considered experimentation with afterimages can be best illustrated using a traditional parallel of human visual system with a photographic device. A moving photo camera gets changing views of surroundings. But after a picture is taken and the lens is shut the captured image stays still with the camera. An afterimage leftover in the head with closed eyes resembles a blurred picture taken by a photo camera. So, what should happen to an afterimage when you slightly move your head?

The common sense and the photo camera analogy suggest that any movements of the head should not affect an afterimage secluded in a head with closed eyes. But contrary to these expectations there is a sensation of afterimage detachment. To experience an afterimage take a look at a bright object, close your eyes, and retain a blur replica of this object. When adaptation to darkness takes a long time, for example on awakening, an afterimage could be quite sharp appearing as a lingering visual picture. Then, by gently turning the head you get a feeling that afterimages do not follow its movements. Whatever the anatomical structures are involved, as long as the whole process is localized inside the head feeling the detachment of afterimages seems unnatural. Dismissing possible sensory inputs brings a conclusion that an autonomous brain is capable to detect changes in absolute position of the body utilizing an outside influx of information. The sensation of afterimage detachment is very robust. This effect exhibits different detachment patterns in relation to geometrical variations of head relocations.

The detachment of afterimages is not just another curiosity among the rich assortment of optical illusions. It is a genuine physical effect – an indication of the absolute frame of reference. In this respect, experimentations with afterimages can be paralleled with the famous Newton's experiment considering water in a rotating bucket. At a starting moment, water fails to follow the turning bucket because of the property of inertia. In observation of afterimages it appears as if they also resist to a rotational movement. Figuratively speaking, it looks like the "liquid of mental information" in the "bucket of human head" possesses a kind of inertia.

Thus, human brain can serve as a detector of absolute relocations. By experiencing the detachment of afterimages, a person in an isolated chamber can detect a change in absolute positioning of this chamber. It should be noticed that this is a static, not a dynamic effect. The detachment of afterimages occurs due to positional changes, not due to acceleration.

Some people, mistakenly, say that the detachment of afterimages is not a physical effect because the brain has some inherent "physiological ability" to retain in memory the position of an observed object. This "objection" can be discarded considering afterimage effects created endogenously. Close your eyes and slightly pressure on your eyeballs. You will get a sensation of seeing bright and dark spots. As you move your head a pattern of these bright and dark spots will not follow the relocation of your head. This shows that a human brain completely isolated from the surroundings can serve as a sensor of absolute relocations in the physical space. Similar detachment effects are observed with other endogenously created images in the brain, like, for example, in lucid dreams.

The most striking feature of afterimages in association with the capabilities to sense the absolute relocations in space has been presented by R.L. Gregory, J.G. Wallace, and F.W. Campbell, "Changes in the size and shape of visual after-images observed in complete darkness during changes of position in space", Quarterly Journ. of Exper. Psychology, Vol. 11, pp. 54-55, 1959.
The presented effect of afterimage detachment had been observed when the head is moved towards and away from the source that produced the afterimage. Contrary to any reasonable anticipation, afterimage shrinks in the former case and enlarges in the latter.

No rational discussion of this very strange outcome has been found in the literature. In the suggested two-step scheme of visual perception, the incoming imagery at the first step is put in the holographic memory of the Universe, then, at the second step it is projected back to the brain.
The back projection reconstruction of an image by a 3D holography is equivalent to that by the lens. With this scheme the explanation of the observed changes of the size of afterimages is straightforward (see Fig. B). When you move your head forward you get a smaller cross-section of the projection cone, when you move your head away you get a bigger cross-section.

The effect of size variation of afterimages can be pressed forward. What will happen if you move your head not a little bit, but at a significant distance in front at more than a meter? The prediction is remarkable. At first the afterimage will vanish, but as you pass the focus point of the converging cone of rays and move further then the afterimage should reappear, increased and inverted.

A preliminary testing of this scheme was done with an afterimage induced by some round shape figure. As the head with closed eyes was moved towards the source of the afterimage the shining spot in the head begun to shrink and then disappeared. When the motion continued, in about 1.5 m, the shining spot of the afterimage reappeared again. Unfortunately, it comes out too blurred to determine that the picture reappears upside-down.

In general, visual perception has some inconvenience: visual imagery is delivered through the lens of the eye in upside-down form. This inconvenience is not taken too seriously since it can be simply taken care of by internal processing of the brain. In other words, what matters is not how the information is delivered but how it is interpreted. The suggested two-step scheme of visual perception provides a better solution. Visual imagery displayed from extracorporeal memory undergoes double reversal: first, by the lens of the eye, and second, by the holographic projection. As a result, visual imagery is delivered to the neuronal circuitry of the brain in a straight form. The presentation of visual imagery by the eye is also associated with the reversion of depth – farthest objects come closer and vice versa. The holographic projection of visual imagery corrects this depth distortion as well simultaneously with flipping the picture turned by the lens of the eye.

If objective confirmation is preferable, afterimages can be explored with laboratory neuroimaging techniques, like fMRI, PET, and microelectrodes. Thus, using the living matter of the brain it is possible to build an improbable physical device - a sensor of absolute relocations of an isolated system. This result is valid irrespective of the developed theoretical view on the organization of biological information processing. "Theory is a good thing but a good experiment lasts for ever" (P. L. Kapitsa).

## 1.1. Moon Illusion as an effect of astronomical aberration

Being based on an outside projection, human perception can be influenced by the absolute motion of the Earth. Perception of a projected image can be changed as a result of dislocation of the receptive area of the brain. An analogous effect presents astronomical aberration in telescopes: owing to the orbital motion of the Earth an observed position of a star can be displaced. In the suggested scheme of visual perception, such an aberration is revealed in the famous Moon Illusion, "one of the most remarkable and surprising illusions".

Celestial objects sometimes appear larger on the horizon and smaller when nearer to the zenith. This effect is called Moon Illusion since the Moon typically displays it, although the enlargement on the horizon is observed also for the Sun, planets, and some constellations, mainly Orion. In principle, size variations in visual perception should occur also in a small scale for common objects in everyday life. However, contrary to celestial objects common objects do not have fixed referral standards and perceptional variations of size in everyday life pass unnoticed.

A visual image, after having been spread in a holographic form over the physical Universe, is reconstructed by a mechanism of the 3D holography acting similarly to optical lenses. Thus, the size of an object can undergo changes when corresponding reception area of the brain is moved out of focus. The same mechanism is related to shrinkage and enlargement of afterimages in the above described *experimentum crucis* as illustrated in Fig. B.

Aberration in holographic reconstruction of visual imagery is determined by displacements of the brain as a result of combinations of different components in the motion of the Earth: daily rotation, annual revolution, and global drift with the solar system. Holographic mechanism is mostly sensitive to angular dislocations. The horizon position of celestial objects may provide most favorable conditions for enlargement of visual imagery. Favorable conditions for observing horizon enlargements of different celestial bodies change periodically with revolution of the Earth. Variations in size of perception are determined by combinations of angular and linear positioning, so these variations can be enhanced when the direction of observation aligns with the velocity of the global drift of the solar system. As indicated in Appendix A, the solar system is moving towards the Virgo cluster, which corresponds to the month of September on Zodiac circle. The constellation Orion is in approximately opposite position with respect to the Virgo cluster. So, enlargements of the Orion at the horizon are observed on a regular basis and are quite pronounced. Remarkably, the maximum possible size of the Moon at the horizon can appear in the month of September producing a spectacular event of the so-called "Harvest Moon". Variations of the apparent size of the Moon are determined by the changes of its position with respect to the ecliptic plane and, as for eclipses, can be predicted with the renowned astronomical accuracy.

"The scientific study of the moon illusion is as old as science itself" ("The Moon Illusion", Edited by M. Hershenson, Lawrence Erlbaum Associates, Publisher, Hillsdale, New Jersey, 1989). It is wishful thinking to believe that the observed changes in the size of the Moon can be explained by atmospheric refraction. The Moon Illusion is not an optical effect - the big Moon never appears on photographic pictures. There are some theories on psychological origin of Moon Illusion, but none of the explanations is adequate. A scrupulous analysis of the problem presented in the referred book (Hershenson, 1989) ends with an candid conclusion that more research on moon illusion "will be of little value": "A more fruitful approach would be to direct research to fundamental issues in visual space perception. If agreement can be reached about those issues, an understanding of the moon illusion would probably be self-evident".

## 2. A new computational model: subsets manipulations in a content-addressable memory without resolution of multiple responses

Characteristics of a computational model include the method of access to the storage, the rules for transformations of information items, the directive how to get the first instruction, and the way how to obtain the next one. Computational process in conventional von Neumann's model is driven by a sequence of execution instructions, which prescribe a particular action and indicate how the next instruction can be obtained. The most prominent property of von Neumann's model is that creating execution instructions presents a part of the computational process. Generally, von Neumann's scheme runs as follows:

(0) Setup an initial pointer to the list of instructions
(1) Fetch a machine instruction
(2) Take information items specified by the instruction
(3) Perform the indicated transformation
(4) Write down the result in the memory
(5) Determine the next instruction
(6) Go to (1)

The von Neumann's scheme is very delicate, sensitive to errors, and non-uniform. Also, it has difficulties in accommodating interrupts for multitasking. Thus, such an organization is not suitable for the control of biological objects.

The new computational model using subset manipulations in a content-addressable memory without resolution of multiple responses produces an unusual framework for the realization of complex control systems.

The key point in effective implementation of computational processes is ease of functionality. The suggested computational model relies on a single universal operational principle: specification and transformation of subsets. The computational process is driven by the information contents. This means that instructions and are data merged in a single entity and special formation of execution instructions is not needed.

Any algorithm can be implemented by a sequence of cycles: Apply a Searching Criterion - Retrieve a Subset and Modify the Searching Criterion. The process is robust and fault-tolerant because it is not sensitive to slight distortions in data-instructions. Multitasking can be naturally implemented thanks to the "atomic" makeup of operational cycles. In biological information processing the suggested computational model comes across a unique situation, which is contrary to usual technological conditions – this model has to operate on slow elements (biochemical structures and neuronal circuits) and very fast memory (holographic facilities of the Universe). Basically, there are two subtypes of this model depending on how the selected subsets in the memory are handled.

The engineering problem of resolution of multiple responses in content-addressable memory has an interesting connection with the mathematical concept of the Axiom of Choice (S. Berkovich and Y. Kochin, Associative Memory, Publishing House "Znanie", Moscow, 1976 (in Russian)). In 1904 E. Zermelo noticed that many mathematical constructions are based on an implicit assumption that there exists a "selection" function that allows indicating a representative element for a given subset. This assumption necessitates a conclusion that elements of any set can be ordered. Formally, it can be said that a mechanism for resolution of multiple responses fulfils the requirements of the Axiom of Choice to enable operations in content-addressable memory.

Operations in the content-addressable memory without a mechanism for resolution of multiple responses can be performed provided that information items have been somehow loaded in the memory beforehand. Loading information items is done on individual basis and thus requires an access to separate memory locations. The argumentation of the Axiom of Choice imposes on the locations of holographic storage a requirement of total ordering. Therefore, the content-addressable memory of the holographic storage must incorporate an addressing system with random or sequential access. This seemingly trivial statement has extensive consequences.

In the context of human memory, the question of how information items can be loaded in a content-addressable storage of the holographic mechanism has not been given a proper consideration. Thus, the necessity for all-encompassing addressing system for human memory has skipped the attention of brain researchers. In the suggested extracorporeal organization of human memory the addressing mechanism of the holographic storage in the physical Universe is provided to biological objects by changes of their angular positioning with respect to the reference wave. These changes continuously occur thanks to the rotation of the Earth.

Interestingly, the suggested model makes the rotational movement of the Earth an engineering requirement for the design. Thus, the addressing scheme in biological memory is implemented as for an electromechanical device, like, for example, for a magnetic tape. Let us emphasize again that the indispensable requirement for an addressing scheme should not be ignored in research models of human memory.

## 2.1 Two forms of biological information processing – inanimate and animate matter

In the first place, the computational model for biological information processing demarcates the division between the dead and living matter. But this model also accounts for drastic differences inside the living matter itself corresponding to the behavior of inanimate and animate matter.

The suggested new type of computational model produces two divergent styles of information processing, which are determined by different approaches to computations in a content-addressable memory without resolution of multiple responses. Subsets taken as a whole can be involved in the following modes of operation: (1) assessment of the cardinality of the retrieved subsets and (2) ranking the elements of retrieved subsets on the basis of similarity with given patterns.

These two modes of operations of the computational model with content-addressable memory present two different tracks in biological information processing. The mode 1 of operations is related to plain control of life functions of organisms – the characteristic behavior of inanimate matter. The mode 2 of operations can perform sophisticated mental activities – the characteristic feature of animate matter.

The general scheme of mode 1 operations – inanimate matter:

    (0)  Set up an initial state
    (1)  Generate an access criterion in conjunction with ongoing biochemical activity
    (2)  Select a subset of information items
    (3)  Evaluate the magnitude of a feedback signal from the selected subset
    (4)  If the magnitude of the signal exceeds certain threshold
            then it impacts DNA inducing a biochemical action
    (5)  Write down to the memory the resulting message of biochemical activity
    (6)  Go to 1

The general scheme of mode 2 operations – animate matter:

    (0)  Setup an initial state
    (1)  Generate an access criterion from available information items
    (2)  Select a subset of information items
    (3)  Rank the obtained information items
    (4)  Modify some top information items
    (5)  Write down modified information items in the memory
    (5)  Go to 1

## 2.2 Technicalities in the implementation of biological information processing

The extracorporeal organization of biological information processing encounters an unusual situation of dealing with fast memory and slow processing elements. In conventional computer science, a system with fast memory and slow processing elements does not attract attention as it goes against the technological grain. The suggested computational model is implemented using the infrastructure of the physical Universe as holographic memory. The access cycle to this holographic memory takes a very short time interval about $10^{-11}$ sec. The processing elements in the suggested computational model are associated with chemical reactions of macromolecules in mode 1 and with electrochemical activities of neurocircuits in mode 2. For both cases, characteristic times of corresponding processes are substantially below the time of memory operations - $10^{-11}$ sec.

For the situation of fast memory and slow processing elements the suggested computational model that manipulates with subsets rather than individual elements becomes very appropriate.

Application of the suggested computational model decisively shapes the whole architecture of biological information processing. In content-addressable operations the consecutive states of transformed subsets preserve their closeness. The overwhelming speed of memory over processing elements brings to digital processing a sense of continuity with a flavor of "analog" computing. This organization of operations endows biological information processing with robustness and fault-tolerance.

Biological information processing is set up for the purpose of control. This means that the important part in biological systems constitutes flows of information to and from the environment. Brain signals from different senses has to be integrated in a single operational mechanism. Although these signals arrive in different places they are processed within a coherent whole. For example, a certain scent may immediately recall a particular visual image. Such an organization becomes possible owing to communication through working memory. Namely, information from the senses does not go directly to the processing elements of the brain, but first goes to the memory. The external signals from the senses do not interrupt the internal process since they deliver information through direct memory access. Thus, sensory inputs are combined in the memory and information from all of the senses becomes available to the brain as a whole. With the fast content-addressable access a system acquires the facilities of central control - the decision making process can take into account all the information in the memory.

The cells of an organism are under global control of the computational model with content-addressable memory in mode 1. Although the scope of yes-no signals of this model is narrow it can give guidance to chemical reactions of macromolecules raising their sophistication to the level of biochemistry. The essential point in this computational scheme is that it may orchestrate the enigmatic process of morphogenesis as an assembly operation in the reference frame of the 3D absolute space. Figuratively speaking, morphogenesis utilizes the GPS of the physical Universe. A system of bacteria with the same DNA presents a singular organism under a global control of the suggested computational model in mode 1.

The different areas of the brain are under global control of the computational model with content-addressable memory in mode 2. The mode 2 computational scheme provides much more sophisticated capabilities. The remarkable feature of this scheme is that it operates with information items on top of ranked subsets. As a result, a large part of information somehow affects the processing activities but do not surface at the output. This feature of the suggested organization of information processing in the brain explains the dominant role of the Freudian mechanism of unconscious.

Taken in isolation, an individual biological object can be seen as a "virtual machine" using a slice of information processing resources of the Universe. The phenomenon of Life is a collective effect where "virtual machines" of different biological objects interact through the common memory of the whole physical Universe. "Virtual machines" of different biological objects are to a great extent independent. But, information accumulated in the Universe in relation to the life cycle of one biological object may influence succeeding biological objects. As a result, evolutionary changes become the major factor in the development of populations of biological objects and present inalienable fact of their existence. The unavoidable changes in populations of biological objects determined by the suggested organization of biological information processing appear as a Lamarkian type evolution since these changes come from accumulation of acquired characteristics. Changes in biological objects may appear as spontaneous events when their actual cause – fixing some information in the infrastructure of the physical Universe – had occurred remotely in space and time. The Darwinian mechanism of evolution – survival of the most reproductively successful – is universally applicable, but it is less efficacious.

## 2.3 Engineering assessment

The suggested computational model shows a number of remarkable engineering characteristics from the standpoint of biological information processing.

(1) Simplicity of software

The major challenge in organization of computations, especially in distributed real-time environment, is managing software complexity. The computational model with contents-addressable memory does not differentiate between instructions and data. Accordingly it does not need complex software. The behavior of this model is simply driven by atomic actions of access and transformation of retrieved subsets for both procedural and operational software. These atomic actions can be naturally intermingled to simultaneously carry out a multitude of different computational activities.

(2) Fault-tolerance

Implementing an ordinary computer system that can tolerate structural distortions is an extremely hard problem. Coping with this problem involves tremendous component redundancy and operational sophistication. On the other hand, the computational model with content-addressable memory naturally provides fault-tolerance to its operations. The universal algorithmic operation of subset manipulation is intrinsically robust as it can sustain distortions in formation of a searching criterion and in evaluation of signals from retrieved subsets.

(3) Reliability

Biological systems show an astounding reliability in retaining their information. Imprints of memory in the brain and immune system can be stored over many decades without a substantial degradation. The spectacular reliability and durability of biological information takes place because its storage is spread over the network in the infrastructure underlying the material world. For a bulk storage based on volatile molecular configurations to have a high level of reliability and long durability would not be possible.

(4) Continual re-writings in read-only memory

The characteristic trait of biological memory is that it is not erasable. The operative reason for this state of affairs is that with the content-addressable access it is not possible to reach the specified location exactly to perform selective writing. Therefore, a memory slice for a particular biological object works with only one permissible write-in operation and unrestricted reading. This gives a specific flavor to the style of algorithmic procedures in biological information processing. Thus, for an information item in a content-addressable storage to attain a higher rank in retrieved subsets it is favorable to have replicated copies of this information item which appear from its multiple re-writings. As a result, the more frequently accessed information items get a higher chance for retrieval.

Thus, the process of continual re-writing gives to the information retrieval in human brain a ranking metric based on the frequency of access. It is worthwhile to note that ranking strategy using the frequency of access has been implemented in one of the most successful "Internet" searching engines – Google.

## 3. Extracorporeal biological information processing in the context of natural science

The conventional scientific paradigm portraying Life as a self-organization process that emerges from randomly scattered molecules is supposed to embrace this phenomenon in its entirety. Thus, the conventional scientific paradigm is loaded with a tremendous burden: it must be able to refer to every finest detail in the behavior of living systems without exception.

Understanding of the phenomenon of Life encounters two layers of problems. First, it is necessary to reveal the operational scheme behind the organization of high performance information processing in living systems. Second, it is necessary to address the meaning of feelings, like pain, pleasure, emotions, and consciousness. The conventional scientific paradigm takes for granted two assumptions: (1) the biological processes are completely controlled by information associated with macromolecule configurations and (2) all the feelings in living systems builds up on intensive information processing.

The suggested organization of biological information processing has a decisive epistemological advantage: it can be absolutely disassociated from the problem of feelings. The suggested organization is aimed only at the engineering design of biological information control systems. The issues of pleasure, pain, emotions, ego, and consciousness can be entirely taken outside of the presented concept. With the suggested concept the question how these lofty issues should be treated can be left open.

However, this is not the case for the conventional scientific paradigm. From the standpoint of the conventional scientific paradigm all the problems in the world, including the problem of feelings, must be completely within the reach of material processes in the physical Universe. Thus, modern "artificial life" projects are forced to believe that a system with intense information processing eventually must build up an assortment of feelings, like joy, love, ambitions etc. Therefore, to keep up with the conventional scientific paradigm the synthetic ability of high performance information processing systems to exhibit the whole plethora of feelings cannot help being taken seriously.

The difference between artificial and natural intelligence is supposed to be established by administering the famous Turing test. A human examiner in an information contact with a remote system has to determine whether this system presents artificial or natural intelligence. In a specified period of time the examiner has to conclude on the status of the assessed system. If no conclusion is achieved, it has to be admitted that such a system is on par with natural intelligence. However, the Turing test does not always produce the desired outcome. For example, the advancements in chess programming give artificial systems an edge over human players.

With the suggested paradigm of extracorporeal organization of biological information processing the difference between artificial and natural intelligence can be established by means of a certain modification to the Turing test. Actually, this modification allows to draw a distinction between information processing performed by living or dead matter. This modification goes along with the *experimentum crucis* for the suggested concept. Suppose a system under examination is placed in an isolated chamber in complete darkness and that this system can be slightly relocated under the control of the examiner. An examiner can request the system to get an image from a flash of light. Then, the examiner can slightly relocate the system and ask whether anything had happened to the recorded image. According to the suggested paradigm of extracorporeal organization of biological information processing a negative answer would reveal an artificial system.

## 4. Annotated bibliography

Vaidyanath, K.S. Jensen, and S. Hameroff, "Computational connectionism within neurons: model of cytoskeletal automata subserving neural networks", Physica D, **42**, 428-449, 1990).

The idea that holographic processing of the brain is an extracorporeal activity in the informational infrastructure of the physical world has been suggested in S. Berkovich, "On the Information Processing Capabilities of the Brain: Shifting the Paradigm", Nanobiology, **2**, 99-107, 1993. This suggestion simultaneously overcomes four barriers in the holographic organization of the brain: (1) informational infrastructure provides very high speed of processing, (2) memory capacity of this infrastructure is virtually unlimited, (3) the structure of the Universe is uniform and unbounded, and (4) the presented CAETERIS model provides continuous generation of coherent reference waves.

2. One of the fundamental questions in the nature of things is why the space we live in has three dimensions and why are we able to perceive only three dimensional objects. These seemingly unrelated questions find a unified resolution in the suggested concept. According to Poincaré the property of three-dimensionality of the space of perception is physiologically determined (H. Poincaré, "Pourquoi l'espace a trois dimensions", Dernieres Pensees, Flamarion, Paris, 1913). But how is this related to the three-dimensionality of the physical space?

In considering wave processes, an important role is played by the Huygens principle. In simple words, this principle presents a rule for the wavefront construction. In abstract form, the Huygens principle is a distinctive property of hyperbolic differential equations with partial derivatives describing the spread of excitation. The case of 3D space has a unique feature that provides spreading of an excitation with a distinct front in a most strict form. Thus, the transmission of signals with high accuracy becomes possible. This is one of the main physical notions brought into play to explain the three-dimensionality of the material world. It is interesting to note that according to a renowned hypothesis by J. Hadamard there exist no other equations with this property.

A storage device with holographic processing is most effective if it employs a wave process operating in accordance with the Huygens principle. The spread of excitation localized within a narrow front is essential for the following reasons. First, it is economic because less elements are activated. Second, fixation of information in the storage medium at intersection of spreading excited regions is more compact. And third, since excited regions occupy a tiny part of the whole storage medium simultaneous processing of many signals may take place.

From these considerations it has been concluded that a model of the brain based on holographic processing cannot effectively provide perception of objects with more than three dimensions: S. Berkovich, "The dimensionality of the informational structures in the space of perception (posing of problem)", Biophysics, **21**, 1136-1140, 1976. At that time, it was believed, however, that the holographic mechanism can been attributed to neurophysiological processes.

In the CAETERIS model the informational infrastructure of the physical Universe serves as a holographic medium for biological information processing. Both, the material and informational processes benefit from having wave propagation conforming to the Huygens principle. This demands for the three-dimensionality of interconnections of the cellular automaton infrastructure of the physical world and makes the global geometry of the Universe as 3D hypersurface of 4D sphere a necessity.

3.  A computational model with distributed associative processing has been presented in:
S. Berkovich, "A New Computational Model for Massive Parallelism", Proceedings of Frontiers '90 – The Third Symposium on the Frontiers of Massively Parallel Computation, IEEE Computer Society, pp. 244-250, 1990.
This computational model is built around a global associative memory with multiplicity of input points and includes resolution of multiple responses through a communication protocol.
A computational model of this type without resolution of multiple responses has been introduced at 1997 International Conference on Computational Physics:
S. Berkovich, E. Berkovich, and G. Lapir, "Fast associative memory + slow neural circuitry = the computational model of the brain", Bulletin of American Physical Society, **42**, No. 6, p. 1575, 1997.

4.  The role of the DNA as a key to shared information processing resources of the physical Universe has been presented in:
S. Berkovich, "The meaning of DNA information in the phenomenon of Life", The Centennial Meeting of the American Physical Society, Atlanta, GA, March 1999, Bulletin of American Physical Society, **44**, No. 1, Part 1, p.115, 1999
(see also http://www.aps.org/meet/CENT99/vpr/laybc31-02.html and APS News, Volume 8, No. 6, p. 3   - "Computing with DNA")

A more detailed analysis of this topic, including consideration of two types of computational models corresponding to inanimate and animate objects, is given in:
S. Berkovich, "On the difference between dead and living matter: making sense of pseudo-random sequences of DNA nucleotides", *The Noetic Journal*, Vol. 2, No 1, pp. 42-51, January 1999.

5.  In the work:  S. Berkovich, "Probing the Architecture of the Brain in Experimentation with Afterimages", Proceedings of the International Joint Conference on Neural Networks Washington, DC, July 10-16, 1999, Volume 1, pp. 69-73,  reconfigurations of afterimages in the head with closed eyes are treated as a geometric optics effect determined by image reconstruction properties in the 3D holography.

The relocations of the head can be classified according to its three basic types of movements with respect to the perceived incidence plane of afterimages: complanar, non-complanar, and orthogonal. The non-complanar relocations occur in rotations of the head about axes parallel to the plane of afterimages, such as in nodding or turning around. In this case, one feels the detachment of afterimages. The complanar relocations are translations in the plane of afterimages and rotations around the axis orthogonal to this plane. They can be implemented with a stiff body in such movements as standing up, sitting down, walking sideward, or tilting. In this case, afterimages do follow the head as if the detachment does not occur. Note that switching from detachment to non-detachment is determined just by supposedly insignificant changes in head rotation whereas in plain kinematical sense nodding and turning look quite similar to tilting.

As reported in (R.L. Gregory, J.G. Wallace, and F.W. Campbell, "Changes in the size and shape of visual after-images observed in complete darkness during changes of position in space", Quarterly Journ. of Exper. Psychology, Vol. 11,  pp. 54-55, 1959) one of the most striking properties of afterimages appears in orthogonal relocations - direct movements towards and away from the source of afterimage. In this case, the afterimages not only do not stay intact but contrary to any reasonable anticipations they shrink in the former case and enlarge in the latter. This feature in afterimage behavior is used for the arrangement of the *experimentum crucis* (see Fig. B).

6. Explanation of Moon Illusion - observed differences in sizes of celestial objects at the horizon and in the zenith - runs into a lot of controversies. "The scientific study of moon illusion is as old as science itself. Both originated during the period 600 to 300 BC, when philosophers in ancient Greece began to propose natural, rather than supernatural, explanations of the world. As the illusion has been studied from the prospective of several disciplines over the centuries, its history reflects the development of the scientific world view. The names of many prominent figures in the history of science feature also in the history of moon illusion" ("The Moon Illusion", Edited by M. Hershenson, Lawrence Erlbaum Associates, Publisher, Hillsdale, New Jersey, 1989).

To distill the astronomical determinant in the Moon Illusion, we have undertaken a Moon watching project "Guarda che Luna". In this project, the attention was shifted from customary consideration of horizon-zenith changes to comparison of variations of the size of the Moon at the horizon in a succession of months. Evidently, variations in the apparent size of the Moon at the horizon observed from the same place but at different times could not be attributed to local optical or psychological circumstances.

The project "Guarda che Luna" has been undertaken in three consecutive months – August, September, and October of the year 2000. The results consist of subjective evaluations of the apparent size of the rising full Moon. These results had been obtained from about 50 respondents from the United States, New Zealand, Jordan, Germany, Taiwan, Belarus, and Russia. In that September, however, the effect of a very big "Harvest Moon" did not happen, so in this particular sequence of observations the differences in the apparent size of the Moon from one month to another had not been exposed as impressively as they could be. Nevertheless, a definite conclusion of this project is that the apparent size of the Moon at the horizon at given place of the Earth varies from month to month and these variations occur coherently in different remote regions over the globe.

From the standpoint of the paradigm of the conventional science there is no clue why the observed size of the horizon Moon should synchronously change over the whole globe from one month to another irrespective to particular terrain and atmospheric conditions. Recognizing this fact would imply a connection between human perception and the construction of the physical Universe. So, should the established worldview be changed to accommodate a seemingly insignificant paradox of "Moon Illusion"?

7. Visual perception that displays information from the memory is freed from a cumbersome requirement for an excessive processing of images that are delivered by the eye in upturned form. Being displayed from the memory, visual imagery undergoes an inversion by the holographic mechanism rendering an adequate representation of the outside world forthwith. The fact that no single case of any disorder linked to upside down delivery of visual images has been ever reported suggests that straightening of the images in the visual system is due to physics rather than to physiology.

The holographic reconstruction also inverts the depth relations, i.e. what was closer appears further and vice versa. But the lens of the eye also reverses the depth. This follows from the formula $1/d_1 + 1/d_2 = 1/f$, where $d_1$ and $d_2$ distances from the lens of an object and its image and $f$ is focus distance. Normally, the issue of depth reversal is not addressed in elementary cases considering projections on a flat screen. In the bulk of the eye system, the situation is different, and an arriving images come in depth reverse form. In the two-step process of perception, the eye inverts the image and reverses the depth relationships first, then the holographic reconstruction inverts the image and reverses the depth relationships back. As result the human brain gets the whole image in a right way in the original form.

The depth dependencies in images in the eye can slightly affect the perceived size of the object. The perceived size of an object can change if it is illuminated by collimated light, so the depth dependencies disappear. Such effects have been observed for the imagery in flight simulators: "uncollimated imagery must be magnified by about 15-30% in order to appear the same size as collimated imagery" (B.J. Pierce, "Magnification of Simulated Targets to Compensate for Decreases in Perceived Size", AFRL Technology Horizons, September 2000, pp. 23-24). The issue of the collimated light as it affects the perceived size of objects has been brought to my attention by R. Potter in conjunction with the discussion on the mechanism of Moon Illusion. Note, that all the celestial objects are observed in collimated light. The differences of the size of the imagery in the eye system may occur due to variations of angle of incoming light. These variations may be very small, but subsequently they are amplified by the holographic back projection.

8. "It's time to stop thinking of perception as a process grounded in separate domains of sight, sound, touch, taste, and smell". (B. Bower, "Joined at the Senses. Perception may feast on a sensory stew, not a five-sense buffet", Science News, vol. 160, pp. 204-205, September 29, 2001). There is a "fundamental mystery of how the brain unites separate sensations into multifaceted experiences".

The scheme of human perception through the suggested direct memory access mechanism clearly shows how the brain develops the integrated form of activity. In the case of sensory perception operating through common memory, all the senses are combined in a unified information processing. Processing through common memory naturally includes past experience and thus features conditioned reflexes.

Also, the perception mechanism through access to the memory may allow input signals to arrive from different sensitive areas of human body. Thus, an interesting effect presents "seeing with a tongue" (P. Weiss, "The Seeing Tongue. In-the-mouth electrodes give blind people a feel for vision", Science News, Vol. 160, pp. 140-141, September 1, 2001). Pattern of pulses representing an image from a video-camera stimulates touch-sensitive nerves of the tongue and give a blind person a possibility to see this image. Tongue stimulation, however, isn't the only way to circumvent blindness. One competing approach is to implant microchips directly in the eyes or brain, another way is to convert images into certain soundscapes, which are piped to a blind person's ears.

In the suggested scheme of human perception the problem of integration of senses is resolved automatically.

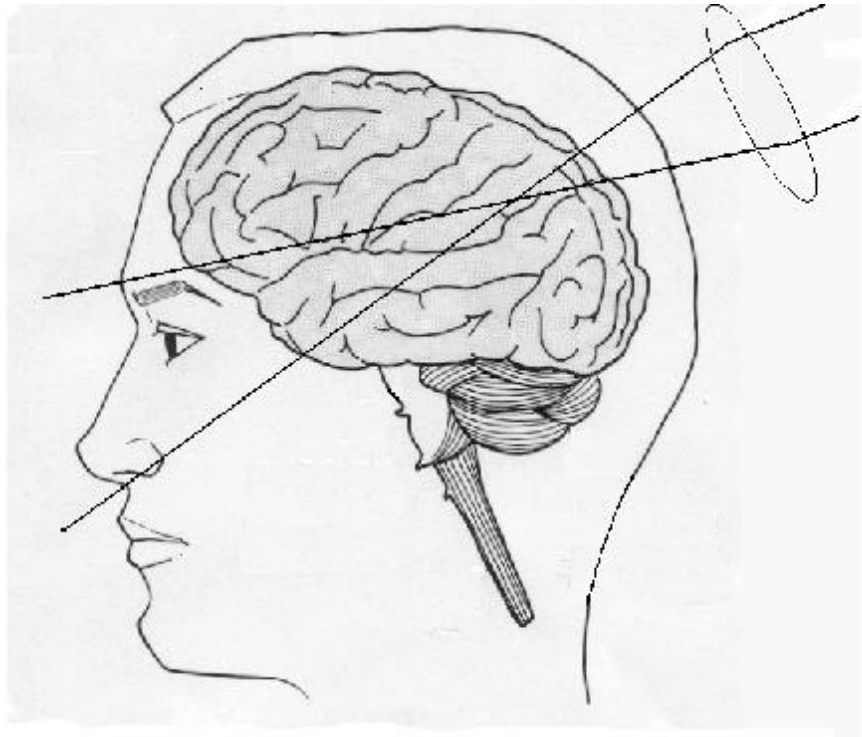

Fig. B

Experiment with afterimage transformation

Visual perception is a two-step process that displays incoming
imagery after it has been fixed in extracorporeal holographic memory.

Projection of images by 3D holography is similar to that by a lens.

This scheme provides an explanation to the incredible behavior of
afterimages in complete darkness reported in (Gregory, 1959):
  "When the head is moved, even by a few centimeters, forward or
  backwards, the after-image changes in size. It increases in size as
  the head is moved back, and decreases as it is moved forward".

The presented scheme implies that if a head with an afterimage
in complete darkness is moved forward faraway, about 1.5 meters,
the decreasing and vanishing afterimage should reappear, in a
bigger size and upside down.